\documentclass[fleqn,usenatbib]{mnras}
\usepackage{float}
\usepackage{newtxtext,newtxmath}

\usepackage[T1]{fontenc}

\DeclareRobustCommand{\VAN}[3]{#2}
\let\VANthebibliography\thebibliography
\def\thebibliography{\DeclareRobustCommand{\VAN}[3]{##3}\VANthebibliography}

\usepackage{graphicx}	
\usepackage{amsmath}	
\usepackage{subcaption}

\title[VHE $\gamma$-ray observations of bright BL Lacs with the LST-1]{VHE $\gamma$-ray observations of bright BL Lacs with the Large-Sized Telescope prototype (LST-1) of the CTAO}

\author[K.~Abe et al]{\parbox{\textwidth}
{\large\centering{
K.~Abe $^{1}$,
S.~Abe $^{2}$,
A.~Abhishek $^{3}$,
F.~Acero $^{4,5}$,
A.~Aguasca-Cabot $^{6}$,
I.~Agudo $^{7}$,
C.~Alispach $^{8}$,
D.~Ambrosino $^{9}$,
F.~Ambrosino $^{10}$,
L.~A.~Antonelli $^{10}$
% ...
(the CTAO-LST collaboration)}}\\\\
Affiliations are listed at the end of the paper
\vspace{10cm}
}

\begin{document}
\label{firstpage}
\pagerange{\pageref{firstpage}--\pageref{lastpage}}
\maketitle

\begin{abstract}
Cherenkov Telescope Array Observatory (CTAO) is the next-generation ground-based $\gamma$-ray observatory operating in the energy range from $20\,\mathrm{GeV}$ up to $300\,\mathrm{TeV}$, with two sites in La Palma (Spain) and Paranal (Chile).
It will consist of telescopes of three sizes, covering different parts of the large energy range.
We report on the performance of Large-Sized Telescope prototype (LST-1) in the detection and characterization of extragalactic $\gamma$-ray sources, with a focus on the reconstructed $\gamma$-ray spectra and variability of classical bright BL Lacertae objects, which were observed during the early commissioning phase of the instrument.
LST-1 data from known bright $\gamma$-ray blazars—Markarian 421, Markarian 501, 1ES 1959+650, 1ES 0647+250, and PG 1553+113—were collected between July 10, 2020, and May 23, 2022, covering a zenith angle range of 4$^\circ$ to 57$^\circ$. The reconstructed light curves were analyzed using a Bayesian block algorithm to distinguish the different activity phases of each blazar. Simultaneous {\it Fermi}-LAT data were utilized to reconstruct the broadband $\gamma$-ray spectra for the sources during each activity phase.
High-level reconstructed data in a format compatible with {\tt gammapy} are provided together with measured light curves and spectral energy distributions (SEDs) for several bright blazars and an interpretation of the observed variability in long and short timescales. Simulations of historical flares are generated to evaluate the sensitivity of LST-1.
This work represents the first milestone in monitoring bright BL Lacertae objects with a CTAO telescope. 
\end{abstract}




\section{Introduction}\label{sec:Introduction}

Blazars are a sub-class of Active Galactic Nuclei (AGN) and constitute the most populous class of sources in the extragalactic very-high-energy (VHE) sky, accounting for nearly 90\% of the known extragalactic VHE source population.
AGN are composed of a supermassive black hole with $10^{6}-10^{9}$ solar masses, surrounded by an accretion disc. The accretion process sometimes powers the formation of ultra-relativistic jets that carry plasmas of high-energy particles.
Blazar jets are closely aligned to the observer's line of sight, and relativistic beaming effects produce strong amplification of the observed non-thermal radiation \citep{urry1995unified}.
Blazars are characterized by rapid variability that spans from radio to VHE $\gamma$ rays ($>100$\,GeV), with timescales ranging from minutes to weeks. However, the ultimate processes driving these emission changes remain largely unknown. Current models explaining variability include: shock acceleration \citep{2022A&A...662A..83D}, turbulence \citep{2019ApJ...887..133B}, magnetic reconnection \citep{2023MNRAS.521L..53A}, jet structure \citep{2012A&A...548A.123B}, accretion changes \citep{2018ApJ...859L..21C}, binary systems \citep{2010A&A...520A..23R}, or even stars crossing the jet \citep{2012ApJ...749..119B}.

Detecting more VHE blazars at different energies, timescales, and distances is crucial for advancing our understanding of their emission mechanisms.
The Spectral Energy Distribution (SED) of blazars typically shows a double-peaked structure. The most common scenario is that the low-energy component—from radio to X-rays—is dominated by synchrotron radiation from relativistic electrons moving in the jet's magnetic field.
In contrast, the origin of the high-energy peak component in the GeV--TeV energy region remains under active discussion in the community.
One widely invoked model---particularly effective for blazars with inefficient accretion flows---is described by the synchrotron self-Compton (SSC) mechanism, in which synchrotron photons produced within the jet undergo inverse-Compton (IC) scattering by the same high-energy electron population \cite[see, e.g.][]{maraschi1992jet}. For blazars with more intense accretion flows, however, SSC is usually subdominant; in these cases, the bulk of the emission is assumed to arise from IC scattering with thermal photons from AGN structures such as the accretion disc, dusty torus, or broad-line region. This process is usually referred to as external-Compton scattering.
As an alternative to these `leptonic' scenarios, various hadronic interactions have been proposed to account for at least part of the emitted high-energy radiation---models that have gained traction partly because they naturally result in the production of high-energy neutrinos in AGN \cite[see, e.g.][]{mannheim1993proton}.

The peak frequency of the low-energy component is used to spectrally classify blazars. While these sub-classifications have evolved over time, and are generally hard to define exactly, here we consider blazars with a synchrotron peak frequency $\nu_s>10^{15}$\,Hz, named high-synchrotron-peaked (HSP).
Intermediate- and Low-frequency-peaked (ISP and LSP) BL Lacs, in contrast, have $10^{14}$\,Hz $< \nu_s<10^{15}$\,Hz and $\nu_s<10^{14}$\,Hz, respectively.
Blazars with $\nu_s>10^{17}$\,Hz are called extreme-high-synchrotron-peaked (EHSP) BL Lacs. This division seems to align with a luminosity gradient in the so-called blazar sequence \citep{fossati1998unifying, donato2001hard}, with LSPs often being more luminous than HSPs. 

Blazars can also be divided into two groups based on the width and intensity of their optical emission lines: Flat Spectrum Radio Quasars (FSRQ) show strong emission lines with |EW| $>5$\,\r{A}. BL Lacertae (BL Lac) objects, on the other hand, typically show weak or no emission lines in that band. FSRQs are often very luminous and are similar to the LSP and ISP classes, while BL Lacs are a more heterogeneous group. 
Of the 84 known VHE blazars (as detected by 2024), 10 are FSRQs and others are BL Lacs\footnote{TeVCat, \url{http://tevcat.uchicago.edu/}}.
Measuring more VHE blazars is necessary to constrain $\gamma$-ray emission models and perform population studies of them.

The Large-Sized Telescope prototype (LST-1) of the Cherenkov Telescope Array Observatory (CTAO) is located on the Roque de los Muchachos in La Palma, Spain. 
Due to its large photon collection area, it is sensitive to energies down to tens of GeV, making it a well-suited instrument to observe $\gamma$-ray sources such as distant AGN that are affected by the photon-photon interactions with the extragalactic background light (EBL). 
From 2020 to 2022, we accumulated more than 150\,hrs of data of several well-known AGN with redshifts in the range 0.03 to 0.5: Markarian 421, Markarian 501, 1ES 1959+650, 1ES 0647+250, PG 1553+113. 

In this work, we present the results on the spectral variability from early observations of bright AGN with LST-1.
We perform a time-resolved spectral analysis using a Bayesian block algorithm for the brightest sources (Markarian 421, Markarian 501, 1ES 1959+650).
To characterize the broadband spectral energy distribution of the high-energy component of each blazar, a contemporaneous joint analysis of {\it Fermi}-LAT and LST-1 data is performed, covering the energy range from 300\,MeV to 10\,TeV. 

\section{Observed Blazars}

\subsection{Markarian 421}

Markarian 421 (often abbreviated as Mrk 421 or Mkn 421) is a high-synchrotron-peaked (HSP) blazar at redshift $z=0.031$ \citep{ulrich1975nonthermal,abdo:2011}. Discovered as a VHE $\gamma$-ray source by Whipple 10-m \citep{1992Natur.358..477P}, it has been monitored by multiple VHE instruments, including Whipple, HEGRA, CAT, H.E.S.S., MAGIC, VERITAS, FACT, Milagro, Telescope Array, and HAWC. 
Mrk 421 is a highly variable blazar across the electromagnetic spectrum. In VHE $\gamma$ rays, variability scales as low as $\sim10$\,min have been detected, with the integral flux being within a range of $\sim0.2\times$ to $>20\times$ the Crab Nebula flux ($C.U.$ in the following), corresponding to the quiescent states and the brightest flares \citep{abeysekara2020great}.

\subsection{Markarian 501}

Markarian 501 (often abbreviated as Mrk 501 or Mkn 501) is another nearby \citep[$z = 0.034$,][]{ulrich1975nonthermal} BL Lac object that has also been the target of multiple Cherenkov astronomy and multi-wavelength astrophysical studies due to its dynamic nature. Notably, it has experienced spectacular flares, such as the 1997 event \citep{aharonian1999mrk501,2000ApJ...532..302A}, making it one of the brightest sources of VHE $\gamma$ rays, with a recorded flux during the maximum of $\sim10$ $C.U.$ in the 
VHE energy range. These historical flares have been key in advancing our understanding of Mrk 501's extreme spectral variability, which behaves as an intermittent EHSP  BL Lac during some of such events, while usually it classifies as a HSP BL Lac \citep{2010ApJ...716...30A, Mrk501_MAGIC_2012}. 

\subsection{1ES 1959+650}

1ES 1959+650 is a bright HSP blazar at a comparatively larger distance than the two Markarians, $z = 0.048$ \citep{perlman1996einstein,2006ApJ...639..761A}, but which has similarly exhibited flares during which the source becomes the brightest source in the $\gamma$-ray sky. It was initially discovered by Telescope Array in 1999 \citep[]{1es1959_TA_1999}, and a major event from it was detected in 2002 by both HEGRA \citep[][reaching a maximum of $\sim2\,C.U.$]{2003A&A...406L...9A} and Whipple \citep[][which reported a peak flux of $\sim5\,C.U.$]{2003ApJ...583L...9H}. In particular, this event provided some of the first evidence of an orphan flare, an extremely bright $\gamma$-ray event without a significant counterpart in X-rays and optical \citep{2004ApJ...601..151K}, and challenging one-zone SSC models that were previously used to explain the broadband spectral energy distribution of the source. Recent flaring events observed by MAGIC, such as the ones on July 1, 2016, have unveiled rapid VHE $\gamma$-ray variability on timescales of minutes, indicating the presence of compact emission regions within its relativistic jet \citep[]{2020A&A...638A..14M} and favouring pure leptonic emission models over hadronic and lepto-hadronic models. 

\subsection{1ES 0647+250}

1ES 0647+250 is another HSP BL Lac previously studied in the VHE regime \citep{2023A&A...670A..49M}. Like with many other BL Lacs, the spectroscopic redshift measurements for 1ES 0647+250 are challenging due to the almost featureless optical spectrum, which is dominated by the continuum. The last estimates are based either on statistical methods using $\gamma$-ray data \citep[$z = 0.45 \pm 0.05$, ][]{2023A&A...670A..49M}, or based on the detection of the host galaxy in the infrared band \citep[$z=0.41 \pm 0.06$, ][]{2011A&A...534L...2K}. Since its early discovery as a VHE emitter in 2011 by MAGIC \citep{aleksic2011gamma}, both MAGIC and VERITAS have observed the source during quiescent states (with fluxes of $\sim3\% \,C.U.$) and during flares reaching fluxes up to $15\% \,C.U.$ \citep{2023A&A...670A..49M,dumm2013highlights}. 

\subsection{PG~1553+113}

PG 1553+113, also classified as a HSP BL Lac \citep{2024MNRAS.529.3894M}, is possibly the most distant source considered in this work. Yet, like 1ES 0647+250, a firm, clear, and unquestioned redshift determination is still lacking. Recent works constrain the redshift in the range $0.433 \leq z < 0.5_{1\sigma,stat}$ \citep{2010ApJ...720..976D,2022MNRAS.509.4330D}, with the lower bound based on Lyman-$\alpha$ forest absorbers up to $z=0.433$ and the no detection of $Ly-\beta$ absorbers, and the upper limit statisticallly constrained by the expected $\gamma$-ray attenuation from EBL. These constraints agree well with other independent, indirect, and statistical estimations based on the expected EBL absorption on the VHE spectrum of the source \citep{2015ApJ...799....7A,2015ApJ...802...65A}. Several remarkable aspects make this source an excellent target for VHE instruments: i) It is well visible from the H.E.S.S., VERITAS and the MAGIC/LST-1 sites, and has been detected by all four instruments \citep{aharonian2006evidence,aliu2015veritas,albert2006detection}; ii) it is a notably bright blazar despite its relatively large redshift, detectable on single nights by the current generation of Cherenkov telescopes; iii) it is one of the few blazars for which multi-wavelength periodicity has been hinted on its long-term light curves \citep{2015ApJ...813L..41A, 2019MNRAS.482.1270C}, interpreted as an indication of a binary supermassive black hole system in the center of the AGN \citep{2018ApJ...854...11T} and the precession effect of its two jets \citep{2021ApJ...922..222H}, or alternatively by pulsational accretion flow instabilities \citep{2020A&A...634A..87L}.

\section{Instruments and Data analysis}\label{sec:Instruments}

\subsection{The Large-Sized Telescope prototype (LST-1)}

The prototype for the Large-Sized Telescope (LST-1) of the Cherenkov Telescope Array Observatory (CTAO), located in the Canary Island of La Palma (28°45'42"N 17°53'30"W) is designed to detect $\gamma$ rays with energies ranging from a few tens of GeV to several TeV. LST-1 boasts a 23-meter diameter dish, and it is a crucial element of the future CTAO-North site, providing access to the lowest energies accessible with the Imaging Atmospheric Cherenkov Telescope (IACT) technique.

LST-1 has been observing the $\gamma$-ray sky regularly since November 2019, and by 2024, it has accumulated more than $2500\,$hrs of observations.

\subsubsection{Data selection}

This work focuses on high-quality data from observations conducted by LST-1 between July 10, 2020, and May 23, 2022, with zenith distance (ZD) ranging from $4\,^{\circ}$ to $50\,^{\circ}$. 
The range is extended up to $57\,^{\circ}$ for 1ES 1959+650 because the data from this source were taken with relatively higher ZD ($36\,^{\circ}$ - $57\,^{\circ}$) and 12\% of the data are above $50\,^{\circ}$.
To ensure consistency and reliability across the datasets, we applied a homogeneous data selection criterion characterized by fixed-quality cuts. Having a systematic and reproducible set of quality cuts was important due to the variability in data acquisition conditions during the early commissioning phase of the LST-1 telescope.

The data were collected in runs, during which the pointing and telescope conditions (night sky background (NSB), weather, camera settings) are roughly stable.
To avoid complications during the reconstruction arising from the increased NSB and correspondingly higher noise in the images, we restricted the analysis to data taken during dark conditions (i.e., no strong Moonlight). 
In line with contemporary practices in Cherenkov telescope observations, most of the data from LST-1 were collected using the Wobble mode \citep{1994APh.....2..137F}. Given the early stage of LST-1 at the time, aspects such as pointing offsets varied from night to night, and issues like tracking stability occasionally arose. To address these issues, we specifically selected observations with stable pointing offsets ranging from 0.35 to 0.45$\,^{\circ}$ from the actual position of the source.

In addition to filtering over pointing offsets, we implemented several other quality checks to ensure the integrity of the data \citep[]{2023ApJ...956...80A}. Pedestal charge stability was assessed by means of the standard deviation during the run, and runs with interleaved pedestal events problems were excluded from the analysis. Similarly, we checked for unstable time resolution, pixel charges, or rates of tagged flat-fielding events. Muon images, obtained during the standard data acquisition, provide insights into the performance parameters of the instrument, such as the optical point spread function and the optical throughput \citep{gaug2019using, 2023ApJ...956...80A}. We therefore monitored muon ring parameters to look for anomalies such as sudden drops in the image sizes. As a proxy for atmospheric conditions, we checked for stable cosmic ray rates and roughly constant average pixel rates for pulses with charges exceeding 10 and 30 photoelectrons. This thorough approach helped us filter out runs with inconsistent or unreliable signal measurements, which could result, for instance, from passing clouds.

Out of the data selection, we ended up with 31.6\,hrs from Markarian 421 (35 nights), 39.5\,hrs from Markarian 501 (46 nights), 11.8\,hrs from 1ES~1959+650 (13 nights), 8.2\,hrs from 1ES~0647+250 (4 nights), and 9.9\,hrs from PG~1553+113 (12 nights).

\subsubsection{Analysis}\label{sec:highLv}

Data calibration and shower reconstruction were performed using the standard LST-1 pipeline {\tt lstchain v0.9.12/13} \citep{2022ASPC..532..357L} as described in \citet{2023ApJ...956...80A}. 
Because the image shower development critically depends on the projection of the geomagnetic field (which changes with arrival direction), the MC generation is performed on a grid of zenith and azimuth
values as described in \citet{2023ApJ...956...80A}.
Because we restricted our analysis to dark extragalactic data only, we explicitly performed the analyses using nominal (i.e., dark) NSB simulated showers.

In order to achieve the best possible performance at each energy, we performed energy-dependent efficiency cuts based on a $80\%$ MC simulated $\gamma$-ray containment, the optimized cuts at the time of the early-commissioning phase, on both \textit{gammaness} and the squared angular distance between the reconstructed event direction and the expected position of the source ($\theta^2$) parameters. 

The energy threshold for the analysis is dependent on the zenith angles. As a default it is set as 10$^{1.4}$\,GeV (25.1\,GeV) for observations with zenith angles below 35$^\circ$~\citep[]{2023ApJ...956...80A}.
We evaluated the energy threshold of LST-1 for a higher zenith range using the MC and LST-1's Instrument Response Functions (IRFs). 
We weighted the number of events in the MC to follow a power-law (PWL) spectrum with an index of -2.4, which is the observed value of 1ES 1959+650 in this work (see Table~\ref{tab:Mrk501_LST_Fermi_SED}).
The PWL function is defined below:
\begin{equation}
\frac{dN}{dE} = f_0\left(\frac{E}{E_0}\right)^{-\alpha}.
\end{equation}
The event selection is then applied to the MC dataset using the same cuts as the experimental data, and the maximum values of true energy histograms of the selected events for each ZD were used as the energy thresholds.
The thresholds were calculated to be about 100 GeV and 150 GeV for ZD = $50\,^{\circ}$ and $57\,^{\circ}$, respectively.

For the light curve analysis, we derived the night-wise $\gamma$-ray flux with energies above 100\,GeV (150 GeV for 1ES 1959+650). The energy threshold was set to the values for variability studies of the integral flux to reduce its relative error.

For the spectral analysis, we tested a PWL and a log-parabola (LP) as candidate models for the intrinsic spectrum. 
The LP function is described as follows:
\begin{equation}
\frac{dN}{dE} = f_0\left(\frac{E}{E_0}\right)^{-\alpha-\beta\textrm{log}\left(\frac{E}{E_0}\right)}.
\end{equation}
The spectrum fit range is from 25.1 GeV to 100 TeV.
The reference energy $E_0$ is set to 300\,GeV for all fits to ensure comparability of the results. The $\gamma$-ray flux is absorbed during propagation due to the interaction with the EBL via pair production, therefore the selected spectral function is folded with the EBL model of \citet{dominguez2011extragalactic}.

\subsection{Fermi Large Area Telescope (LAT)}

The Large Area Telescope (LAT) onboard the {\it Fermi} satellite is a pair-conversion telescope sensitive to $\gamma$ rays with energies from $20\,$MeV to more than $300\,$GeV. With an instantaneous field of view covering $20\%$ of the entire sky, {\it Fermi}-LAT is usually operated in survey mode to provide full-sky coverage every $\sim 3$\,hrs \citep{2009ApJ...697.1071A}. These two characteristics make LAT the perfect accompanying instrument for LST-1 observations. First, it extends down the energy range covered by LST-1 to the MeV band. Second, the energies covered by {\it Fermi}-LAT give insights about the intrinsic $\gamma$-ray spectrum as they are usually found in the optically thin regime (both regarding possible intrinsic absorption in the case of FSRQs, and EBL in the case of distant sources). Because LAT is continuously scanning the sky, it also offers a more uniform temporal coverage to characterise possible flaring blazar activity and to contextualize the LST-1 observations. Finally, LAT data is extensively used by the LST collaboration to initiate Target Of Opportunity and follow-up programs on promising flaring events. 

\subsubsection{Data selection and analysis}

{\it Fermi}-LAT Pass 8 SOURCE class events with energies larger than $300\,$MeV were collected in a region of interest of $10\,^{\circ}$ radius around the nominal position of each selected target. The analysis was done using the standard Fermi Science Tools \citep{2019ascl.soft05011F} using two independent high-level analysis wrappers for cross-validation purposes: {\tt enrico} \citep{2015ascl.soft01008S} and {\tt fermipy} \citep{2017ICRC...35..824W}. In both cases, we used the last available version ({\tt P8R3\_SOURCE\_V3}) of LAT's IRFs.
We set a conservative zenith cut of $90\,^{\circ}$ to avoid Earth's limb contamination and setting as an additional filter with {\tt DATA\_QUAL==1 \&\& LAT\_CONFIG==1} as described in Cicerone\footnote{\url{https://fermi.gsfc.nasa.gov/ssc/data/analysis/documentation/Cicerone/Cicerone_Likelihood/Exposure.html}}. The full sky model was generated using as starting point the {\it Fermi}-LAT $12$-Year Point Source Catalog (4FGL-DR3) \citep[]{Abdollahi_2022} adding all sources and their respective catalog spectral parameter values within $20\,^{\circ}$ radius around the position of the source. During the model optimization, we froze all spectral parameters but the normalization for weak ($<10\,\sigma$ significance in the 4FGL) and far-away sources ($>5\,^{\circ}$). For the target source, we tested the same three spectral model shapes as with LST-1, both with and without taking into account EBL absorption: PWL and LP. 
We selected PWL + EBL-based analysis since the test statistic ($TS$) values of other models over PWL + EBL were less than 9 ($\sqrt{TS} < 3$).
The positions (and extension, whether applicable) of all sources in the full sky models were frozen to the catalog values.

\subsection{Joint analysis of LST-1 and {\it Fermi}-LAT data}

To obtain a clearer intrinsic $\gamma$-ray spectrum from the observed data, we adopt a joint-fit spectrum analysis method between LST-1 and {\it Fermi}-LAT.
We use {\tt Asgardpy v0.4.4} \citep{asgardpy_044}\footnote{\url{https://asgardpy.readthedocs.io/en/latest/}}, an extended support tool for {\tt Gammapy v1.1}.
It uses the DL3 format data (specified by the Data Formats for Gamma-Ray Astronomy \citep[]{gadf}) of different instruments, to be analyzed with joint-likelihood fits and creating the DL5 (SED) products. Although {\tt Gammapy} can perform multi-instrument analyses, it is limited in combining data produced directly from the default formats of each instrument. {\tt Asgardpy} performs the additional functions required to perform analysis with {\tt Gammapy}.
For {\it Fermi}-LAT, it supports both {\tt enrico} and {\tt fermipy} file structures, for transforming the files to {\tt Gammapy} readable format.

We limited the analysis of {\it Fermi}-LAT data to the energy range 300\,MeV - 300\,GeV through a safe energy mask in {\tt gammapy}. Similarly, for LST-1 data, we set 25.1\,GeV - 25.1\,TeV for Mrk 421, Mrk 501, and 1ES 1959+650, and 25.1\,GeV - 1\,TeV for the other two sources in order to get approximate $\gamma$-ray spectra and avoid gaps between the instruments.
For each spectrum, the preferred spectral model is calculated using the Log-Likelihood Ratio test (LRT).
We set a reference energy $E_0$ to 300\,GeV as done for the single instrument LST-1 fits.

\section{Results}\label{sec:Results}

\subsection{Analysis approach}

For each variable source in our sample, a Bayesian block analysis based on the algorithm implemented in {\tt astropy} \citep{scargle2013studies} was carried out to separate states of activity and identify time periods with roughly stable emission.
As input for the algorithm, we used the nightly flux light curve of each source reconstructed by LST-1 above 100\,GeV (150\,GeV for 1ES 1959+650), and added the systematic uncertainty 6\% determined in \citet{2023ApJ...956...80A} to the measured fluxes. The false alarm probability of 3$\,\sigma$ ($p$-value = 0.27\%) was then used to define the individual blocks. 
Since only the central time of an observation is given to the algorithm, it can happen that a block edge occurs in the middle of an observation. For these cases, we moved the block edge to the start or end of the observation.
We performed spectral fits for each observation using an LP model and confirmed that the parameters vary within less than 3$\,\sigma$ significance in each block. 
In the second stage, we performed the same spectral fits for each block and compared the spectral parameters between all blocks. We then manually merged those blocks with parameter values compatible within 3$\,\sigma$ significance. We additionally checked that there are no significant disagreements between the spectral shape and SED points of the blocks to be merged.
For each merged block spectrum, PWL or LP fits were applied, and a preferred spectral model was calculated using the LRT. 
The energy threshold of the spectra except for 1ES 1959+650 is set as 10$^{1.4}$\,GeV (25.1\,GeV) for observations with zenith angles below 35$^\circ$, and 100\,GeV with other zenith angles, to cover the range of the whole dataset (4$^\circ$ - 50$^\circ$).
The energy threshold for 1ES 1959+650 is set as 150\,GeV to cover 36$^\circ$ - 57$^\circ$ as described in section~\ref{sec:highLv}.

For each source and Bayesian block, a dedicated {\it Fermi}-LAT data analysis was produced within the temporal limits of each block, with a similar testing and selection of the best-fitting intrinsic spectral model. 
We performed a fit of {\it Fermi}-LAT spectra in each block and checked whether the block-wise spectral parameters in the same amplitude bins vary within 3$\,\sigma$.
In case the spectral parameters vary more than 3$\,\sigma$, the merged blocks are separated again. 
Then the spectra of both instruments are compared in each spectral state. 

\subsection{Long-term light curves}

\begin{figure*}
\centering
\includegraphics[width=0.85\textwidth]{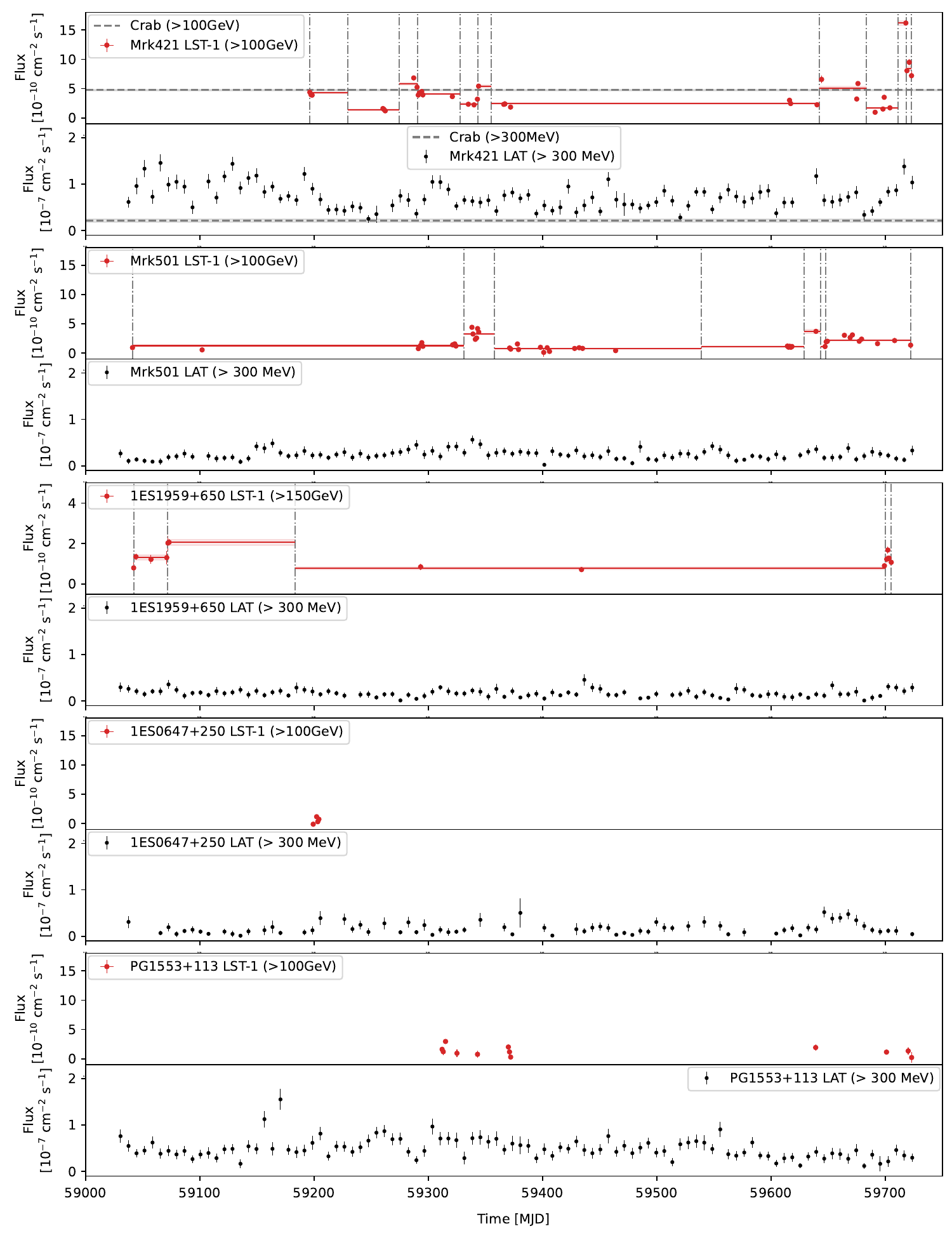}
\caption{LST-1 and {\it Fermi}-LAT flux light curves for energies above $100\,$GeV (150\,GeV for 1ES\,1959+650) and $300\,$MeV, respectively. 
From top to bottom, Mrk 421, Mrk 501, 1ES 1959+650, 1ES 0647+250, and PG 1553+113 are shown. Vertical gray lines show the bin edges obtained from applying the Bayesian block algorithm and the horizontal red lines and shaded areas show the average flux in each block and its standard deviation, respectively.
The Crab Nebula flux is depicted with a horizontal gray dashed line \citep{aleksic2015measurement}.}
\label{fig:Fermi_LST_LC}
\end{figure*}

The LST-1 flux light curve in the whole data period is presented in Fig.~\ref{fig:Fermi_LST_LC}. The Crab Nebula flux in the same energy range is depicted with a horizontal gray dashed line.  
The LST-1 fluxes are mainly less than the Crab Nebula flux except for Mrk 421. The dataset was separated into 11, 7, and 4 blocks for Mrk 421, Mrk 501, and 1ES 1959+650, respectively, by the Bayesian block analysis.
We did not find significant flux variability for the other two sources.

There was a Mrk 421 flaring event (reaching $\sim3.5\,C.U.$) with a duration of about a day on 2022 May 18 (MJD 59717). The Mrk 421 spectral index $\alpha$ with a PWL fit lies between 1.9 and 3.2 depending on the night (see 
Sec. \ref{sec:index_flux}), while the flare spectrum is best described by a LP shape with $\alpha$ = $2.02\pm0.04$ and $\beta$ = $0.10\pm0.02$. 

In Fig.~\ref{fig:Fermi_LST_LC}, we also show the {\it Fermi}-LAT light curves of the observed five sources within the LST-1 observation periods.
The light curves of four sources, except Mrk 421, have moderate flux variability in the LST-1 data time period.

Using the number of Bayesian blocks found as an indicator for the level of variability in the VHE band, Mrk\,421 shows the highest degree of variability, followed by Mrk\,501 and 1ES\,1959+650. The same trend is found, when using the fractional  $F_\text{var}$ as an indicator \citep{2003Vaughan}\footnote{\url{https://docs.gammapy.org/1.3/api/gammapy.stats.compute_fvar.html}} with $F_\text{var, VHE}^\text{Mrk 421}=0.71 \pm 0.02$, with $F_\text{var, VHE}^\text{Mrk 501}=0.61 \pm 0.03$, with $F_\text{var, VHE}^\text{1ES\,1959+650}=0.30 \pm 0.05$. The variability in the \textit{Fermi}-LAT band is systematically lower with $F_\text{var, HE}^\text{Mrk 421}=0.21 \pm 0.05$, $F_\text{var, HE}^\text{Mrk 501}=0.24 \pm 0.05$, with $F_\text{var, HE}^\text{1ES\,1959+650}=0.21 \pm 0.07$.

\subsection{Spectral index vs. flux correlation}\label{sec:index_flux}
\begin{figure}
\centering
\includegraphics[width=0.9\columnwidth]{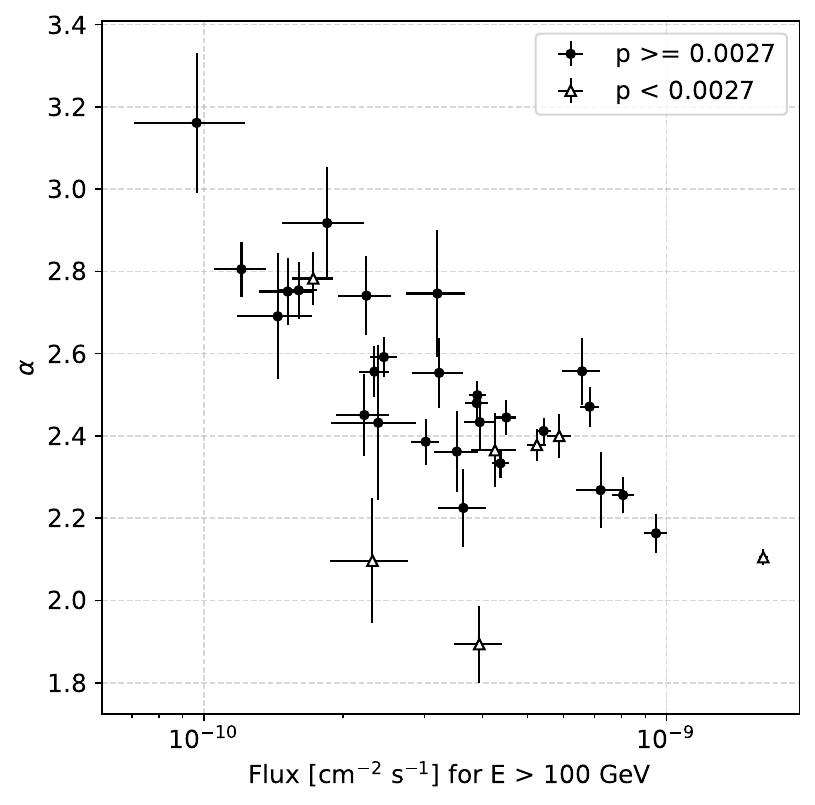}
\caption{Correlation between PWL spectral index $\alpha$ and the night-wise integral flux, as observed with LST-1 above 100\,GeV for Mrk 421.
Filled circles and open triangles indicate spectrum fits that result in a $p$-value above and below 0.27\%, respectively.}
\label{fig:Mrk421_indamp}
\end{figure}

Fig.~\ref{fig:Mrk421_indamp} shows the spectral index as a function of flux for daily measurements of Mrk 421 with LST-1. 
The spectrum fitting was performed with a PWL function since the alternative curved models are not preferred at a significance of 3$\,\sigma$ or more on a nightly basis.
The reference energy $E_0$ is fixed at 396\,GeV, which is determined by the decorrelation energy of a PWL fit with the given reference energy~\citep{abdo2009fermi} using the whole data of Mrk\,421.

The index variation shows a clear harder-when-brighter behavior~\citep{acciari2021multiwavelength}. 
The weighted Pearson coefficient \citep{alexander2013improved} is $-0.81^{+0.04}_{-0.08}$ when one considers only the nights with PWL fits having a significance of 3$\sigma$ or less ($p$-value above 0.27\%).
The discrete correlation function \citep[DCF;][]{edelson1988discrete} shows a correlation coefficient of -0.74 $\pm$ 0.24.
There are no large outliers in the plot, indicating that the $\gamma$-ray emission condition does not change much within the dataset.

The Mrk 501 and 1ES 1959+650 spectral index variation showed no evidence of a strong correlation with the flux within the datasets. The weighted Pearson coefficient and DCF values for Mrk 501 are $-0.18^{+0.19}_{-0.22}$ and -0.54 $\pm$ 0.49, respectively.
The weighted Pearson coefficient for 1ES 1959+650 is $0.24^{+0.36}_{-0.29}$.
The DCF cannot be calculated from the dataset of this source, due to the too small variation of the average index values compared to their errors.

\subsection{Intra-night variability of Mrk 421}\label{sec:intraLC}
\begin{figure}
\centering
\includegraphics[width=\columnwidth]{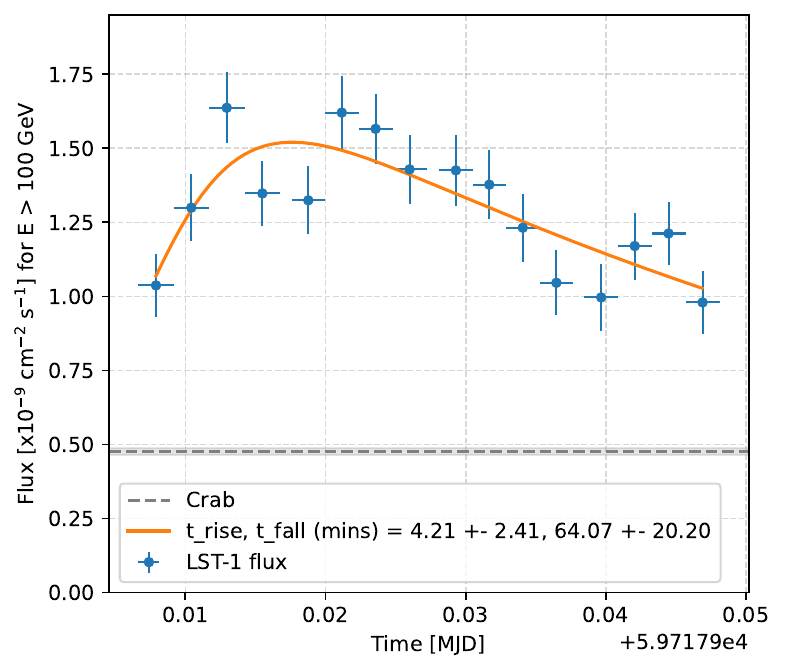}
\caption{Intra-night variability of Mrk 421 flare on MJD 59717. The light curve above 100\,GeV is constructed with 4 data runs and the total exposure time is $\sim53$\,min. A time-binning of $\sim3.3$\,min is used. Note that "run-wise" corresponds to one wobble observation time, which is $\sim13$\,min in this data. The solid orange curve corresponds to a fit with the function given in Eq. \ref{eq:risefall}. Gray shows the Crab Nebula flux above 100\,GeV as in Fig.~\ref{fig:Fermi_LST_LC}.}
\label{fig:Mrk421_intraLC}
\end{figure}

The observed Mrk 421 flux on 2022 May 18 (MJD 59717) exhibited a bright flare with a duration of about a day, with flux at the peak of the flare of 3.5\,$C.U.$ above 100\, GeV. 
We analyzed the light curves with a time-binning of $\sim3.3$\,min to search for short time-scale variability, as depicted in Fig.~\ref{fig:Mrk421_intraLC}. 
For this study, we used {\tt lstchain v0.10.11} and IRF of telescope pointing for each time bin interpolated from the 
MC nodes.
The null hypothesis with a constant flux is rejected at 5.3$\,\sigma$ using an LRT.

We obtained the rise and decay times by fitting the peak in the light curve with the sum of two exponential functions:

\begin{equation}\label{eq:risefall}
F(t) = A_0 / (e^\frac{t_0 - t}{t_r} + e^\frac{t - t_0}{t_f})
\end{equation}

The $A_0$, $t_r$, and $t_f$ correspond to two times the flux at $t_0$, the rise time of the flare, and the decay time of that, respectively.
The flux doubling time is defined as $t_{rise/fall} = t_{r/f} \times ln(2)$ in this formalism.
To investigate a systematic error, we analyzed different cases with time-binning from 2 to 7 min, and compared the average values of $t_{rise}$ and $t_{fall}$.
The rise and decay times are calculated as $t_{rise} = 4.2 \pm 2.4$ (stat.) $\pm$ 1.2 (syst.) min and $t_{fall} = 64 \pm 20$ (stat.) $\pm$ 6 (syst.) min, respectively, with the fit $\chi^2$/dof = 17/12.

Assuming the emitting region is a spherical blob of radius $R$, the size of the radiating blob in the co-moving frame of the jet is constrained by the following relation \citep{tavecchio2010constraining}:

\begin{equation}
R \leq \frac{c\,t\,\delta}{1+z}
\end{equation}

The Doppler factor $\delta$ is related to the bulk Lorentz factor and the viewing angle of the jet. 
$z$ represents the redshift of the source, 0.031 for Mrk 421.
We use the fastest variability timescale $t \sim 5$\,min in this calculation.
Under the assumption of $\delta = 10-50$, which is often used to model HSP-type objects, the upper limit to $R$ is constrained to $(1-5) \times 10^{14}$\,cm.
The mass of the central supermassive black hole in Mrk 421 is estimated as $M = 10^{8.29}\,M_\odot$, where $M_\odot$ denotes one solar mass \citep{wang2004connection}.
Using the gravitational radius $r_g = G_N M / c^2$ ($G_N$: Gravitational constant), the limit to $R$ corresponds to ($3.4 - 17.5$) $\times$ $r_g$.

\begin{table*}
  \caption{EBL-corrected spectral fit parameters of Mrk421 for LST-1 results. The reference energy $E_0$ is fixed at 300\,GeV.}
  \label{tab:Mrk421_LST_Fermi_SED}
  \centering

  
  \begin{tabular}{lccccccccc}
    \hline \hline
    State & Time & PWL fit & & & LP fit & & & & LRT\\
    & [MJD] & $f_0 \times 10^{-10}$& $\alpha$ & $\chi^2$/dof & $f_0 \times 10^{-10}$& $\alpha$ & $\beta$ & $\chi^2$/dof & [$\sigma$]\\
     & & [TeV$^{-1}$ cm$^{-2}$ s$^{-1}$] & & & [TeV$^{-1}$ cm$^{-2}$ s$^{-1}$] & & & & \\
    \hline
    low & 59229.22-59274.68 & 1.96 $\pm$ 0.08 & 2.78 $\pm$ 0.03 & 37.0/18 & 2.12 $\pm$ 0.11 & 2.81 $\pm$ 0.04 & 0.05 $\pm$ 0.03 & 31.8/17 & 2.3\\
     & 59683.52-59710.96 &  &  & & &  & & & \\
    mid-low & 59327.96-59343.44 & 2.60 $\pm$ 0.09 & 2.55 $\pm$ 0.03 & 48.1/18 & 3.09 $\pm$ 0.13 & 2.49 $\pm$ 0.05 & 0.18 $\pm$ 0.04 & 8.32/17 & 6.3\\
     & 59354.96-59642.17 &  &  & & &  & & & \\
    middle & 59196.17-59229.22 & 4.09 $\pm$ 0.09 & 2.44 $\pm$ 0.02 & 114/18 & 4.75 $\pm$ 0.12 & 2.30 $\pm$ 0.04 & 0.18 $\pm$ 0.03 & 14.6/17 & 10.0\\
     & 59290.60-59327.96&  &  & & &  & & & \\
    high & 59274.68-59290.60 & 5.58 $\pm$ 0.12 & 2.44 $\pm$ 0.02 & 92.1/18 & 6.36 $\pm$ 0.16 & 2.33 $\pm$ 0.03 & 0.16 $\pm$ 0.02 & 13.4/17 & 8.8 \\
     & 59343.44-59354.96&  &  & & &  & & & \\
     & 59642.17-59683.52 &  &  & & &  & & & \\
    post-flare & 59718.42-59722.91 & 7.94 $\pm$ 0.31 & 2.26 $\pm$ 0.03 & 42.1/18 & 9.12 $\pm$ 0.40 & 2.10 $\pm$ 0.07 & 0.18 $\pm$ 0.04 & 11.4/17 & 5.5\\
    flare & 59710.96-59718.42 & 14.9 $\pm$ 0.36 & 2.14 $\pm$ 0.02 & 28.2/18 & 15.8 $\pm$ 0.43 & 2.02 $\pm$ 0.04 & 0.10 $\pm$ 0.02 & 5.36/17 & 4.8\\%
    \hline
  \end{tabular}
\end{table*}
\subsection{Spectral analysis}

\subsubsection{Spectral variability}
\label{subsec:spec_var}
\begin{figure}
\centering
\includegraphics[width=0.5\textwidth]{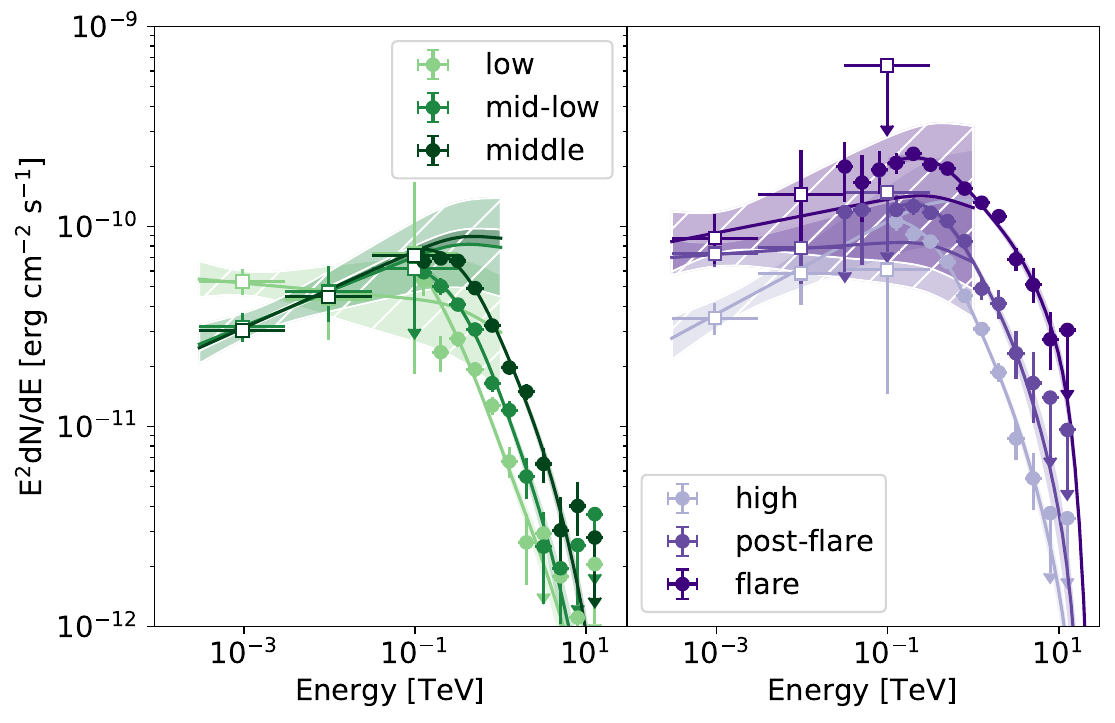}
\caption{Spectral energy distribution of Mrk 421, segmented into six blocks identified by the Bayesian block algorithm, with filled circles representing LST-1 observations and open squares denoting {\it Fermi}-LAT observations. The hatched and shaded areas indicate the statistical uncertainties of the spectral fits for {\it Fermi}-LAT (PWL model) and LST-1 (LP model) including EBL absorption, respectively, with each dataset fitted independently. The corresponding model parameters are shown in Tables~\ref{tab:Mrk421_LST_Fermi_SED} and \ref{tab:fermi_fit}.}
\label{fig:Mrk421_LST_Fermi_SED}
\end{figure}

The $\gamma$-ray spectral energy distributions for the six selected blazars are shown in Fig.~\ref{fig:Mrk421_LST_Fermi_SED} and Fig.~\ref{fig:LST_Fermi_SED}. The best-fit model for LST-1 is represented as a solid color confidence band, with the different colors denoting different Bayesian blocks. Flux points are similarly represented as filled circles. The best fits to the matching {\it Fermi}-LAT datasets are represented as a white-hatched confidence band, and the corresponding flux points are represented as open circles. 

For Mrk 421, 6 states were identified out of 11 Bayesian blocks.
Table~\ref{tab:Mrk421_LST_Fermi_SED} describes the spectral fit parameters.
The LP model is preferred compared to the PWL model with more than 3$\sigma$ for all states.
The index $\alpha$ changes from 2.81 $\pm$ 0.04 to 2.02 $\pm$ 0.04, with a relatively constant $\beta$ change (from 0.05 $\pm$ 0.03 to 0.18 $\pm$ 0.03), where higher flux state showed a harder index as the night-wise flux results (see Fig.~\ref{fig:Mrk421_indamp}).
For Mrk 421, LST-1 and {\it Fermi}-LAT SEDs show good agreement between both results in all spectral states.

For Mrk 501, 6 spectral states were identified from 7 Bayesian blocks. 
The spectral parameters for all 6 states are listed in Table~\ref{tab:Mrk501_LST_Fermi_SED}. The LP model is preferred compared to the PWL model with more than 3$\sigma$ for 4 states.
The index $\alpha$ changes from 2.26 $\pm$ 0.13 to 1.55 $\pm$ 0.03, with a relatively constant $\beta$ change (from 0.11 $\pm$ 0.07 to 0.27 $\pm$ 0.11).

For 1ES 1959+650, 3 states out of 4 blocks were identified.
The LP model is not preferred compared to the PWL model with more than 3$\sigma$ for 2 states. The LST-1 data for the PWL model showed a similar spectral index but a different flux between the states~(see Table~\ref{tab:Mrk501_LST_Fermi_SED}).
LST-1 and {\it Fermi}-LAT spectra are in agreement for middle and high states (see Fig.~\ref{fig:1ES1959+650_LST_Fermi_SED}).
For the low state, the statistics of {\it Fermi}-LAT data simultaneous to the LST-1 observations ($\sim6$\,hrs) are too small to fit the spectrum.
In this case, we adopted a joint-fit analysis method between LST-1 and {\it Fermi}-LAT (explained in Sect. \ref{joint}).

For 1ES 0647+250 and PG 1553+113, the spectral data fits are done over the full time period and shown in Fig.~\ref{fig:1ES0647+250_LST_Fermi_SED}-\ref{fig:PG1553+113_LST_Fermi_SED}.
The data points of both sources show a spectral softening feature at lower energy than around 100\,GeV, where the EBL effect starts.
For 1ES 0647+250, the SED of {\it Fermi}-LAT with the PWL fit does not show a good agreement, even though now there is no preference for an LP fit  (see Table~\ref{tab:Mrk501_LST_Fermi_SED}) found due to limited statistics. We still additionally show the LP fit, which resolves the apparent discrepancy between the instruments.

\subsubsection{Systematic uncertainty}
Systematic uncertainties are tested using the Mrk 421 dataset during the flare state as a benchmark, as its spectrum during that period has the widest energy coverage among all the available data, spanning from 25\,GeV to $\sim10$\,TeV.

\begin{table}
  \caption{Systematic errors of SED model parameters and flux above 100\,GeV considered in this work.}
  \label{tab:syst}
  \centering
  \begin{tabular}{lcccc}
    \hline \hline
    Systematic effect  & $f_0$ & $\alpha$ & $\beta$ & Flux \\
    \hline
    $\gamma$-ray efficiency & 7.5\% & 5.4\% & 24\% & 15\%\\
    \hline
    Background & $<$0.1\% & 0.9\% & 10\% & 1.5\% \\
    \hline
    Day-by-day & & & & 6\% \\
    \hline
  \end{tabular}
\end{table}

The following possible sources of systematic uncertainties are evaluated: 

\textit{$\gamma$-ray efficiency}: 
In order to estimate the systematic uncertainty arising from a non-perfect simulation of $\gamma$-hadron separation efficiency and from errors in the assumed $\gamma$-ray point-spread function~(PSF), we tested the stability of the reconstructed SED against variations in the assumed \textit{gammaness} and $\theta^2$ cut efficiencies. 
We applied combinations of \textit{gammaness} and $\theta$ selection cuts, with (40, 80, 90)\% and (80, 90)\% $\gamma$-efficiency for \textit{gammaness} and $\theta$, respectively.
As a result, the SED model parameter difference is within 7.5\%, 5.4\%, and 24\% for $f_0, \alpha$, and $\beta$, respectively. 
The difference of flux above 100\,GeV is within 15\%.
Note that nominal values of 80\% were selected for the rest of this work.

\textit{Background estimation}: 
As a background-dominated experiment, the reconstruction of LST-1 is sensitive to variations in the background estimation and gradients of it across the field of view (FoV). To quantify this effect, we compared the background rate after event cleaning and selection in two different off regions
at the same distance from the center of the field of view ($0^{\circ}_{\cdot}4$) and Mrk 421 ($\sqrt{2} \times$$0^{\circ}_{\cdot}4$), using the Mrk 421 data with zenith angles below 35$^{\circ}$ (25.5\,hrs in total).
We obtained a difference between them by 0.3\%$\pm$0.2\% for the lowest energy bins considered in this work; between $10^{1.4}$\,GeV and $10^{1.8}$\,GeV (25\,GeV and 63\,GeV).
We analyzed the SED with increased background rates by 0.3\%, and the SED model parameters are changed within $<$0.1\%, 0.9\%, and 10\% for  $f_0, \alpha$, and $\beta$, respectively. 
The difference of flux above 100\,GeV is 1.5\%.
We also tested an analysis method with a different background event sampling method \citep[source-dependent analysis][]{2023ApJ...956...80A}, and the difference from the standard (source-independent) analysis used in this work was within statistical errors \citep{takeishi2023pos}.

\textit{Day-by-day systematics}: 
We also applied a day-by-day flux systematic error to the analyzed light curves for Bayesian block selection.
Systematic uncertainty of 6\% on the nightly flux values is referred from the Crab Nebula light curve study with LST-1 \citep{2023ApJ...956...80A}.

The whole systematic errors considered in this work are summarized in Table \ref{tab:syst}.
$\gamma$-ray efficiency dominates for bright sources, and background stability dominates for weak sources. 
Day-to-day variations in the atmosphere and telescope throughput are relevant when considering the time evolution of a source signal.

\subsection{Joint-fit analysis}\label{joint}

\begin{figure*}
\centering

\begin{subfigure}[t]{0.45\textwidth}
    \centering
    \includegraphics[width=1\textwidth]{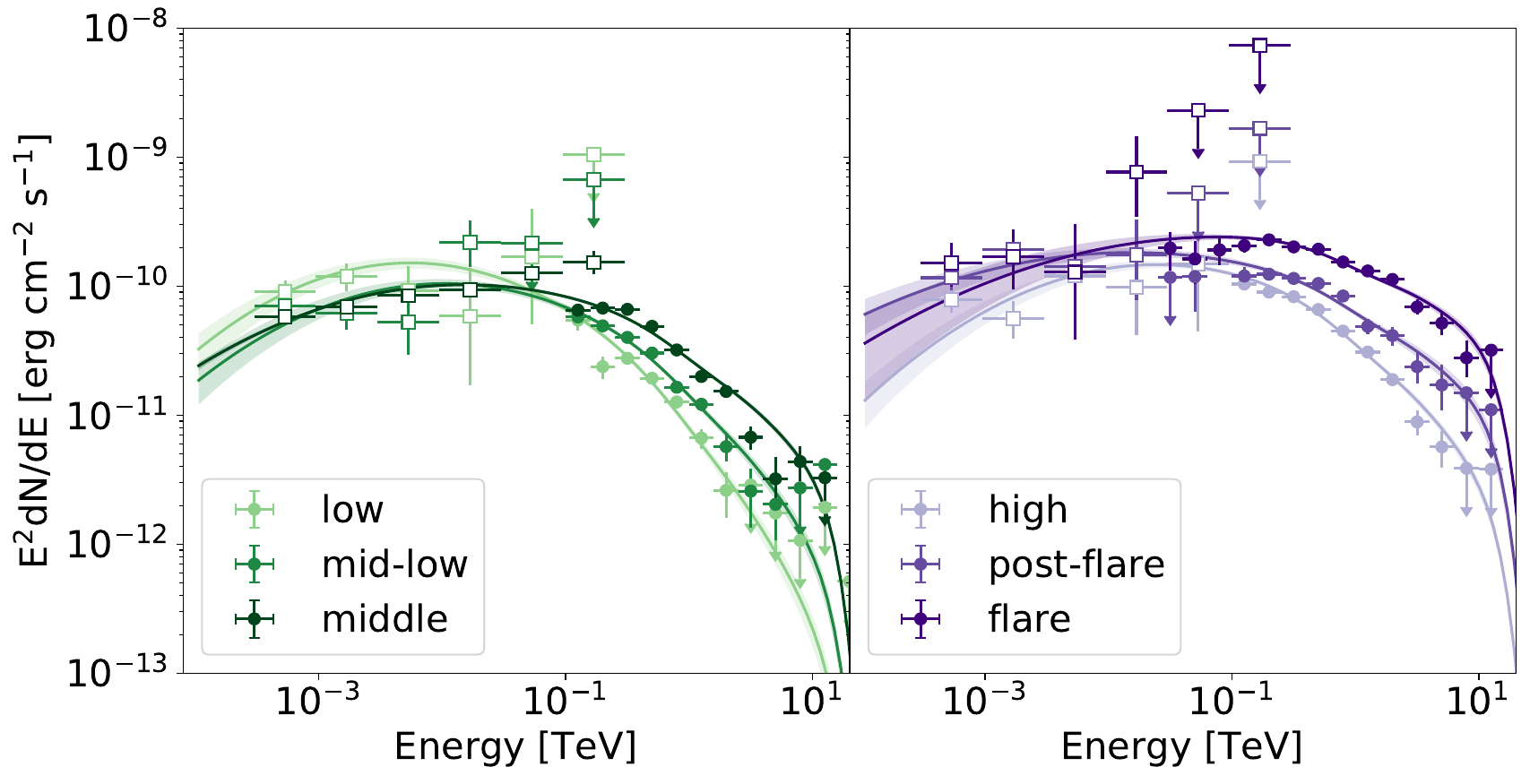}
    \caption{Mrk 421}
    \label{fig:Mrk421_joint}
\end{subfigure}
\begin{subfigure}[t]{0.45\textwidth}
    \centering
    \includegraphics[width=\textwidth]{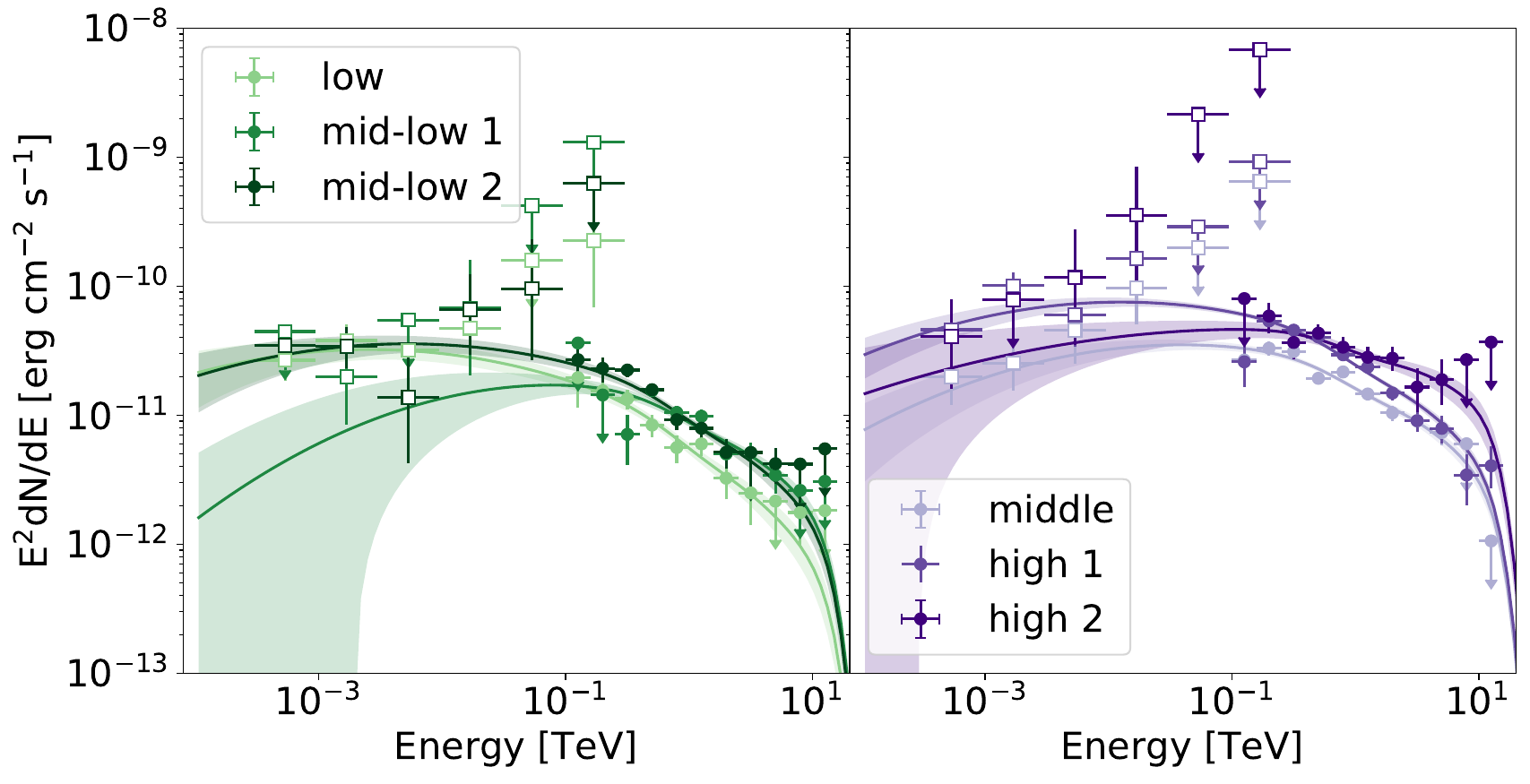}
    \caption{Mrk 501}
    \label{fig:Mrk501_joint}
\end{subfigure}
\begin{subfigure}[t]{0.3\textwidth}
    \centering
    \includegraphics[width=\textwidth]{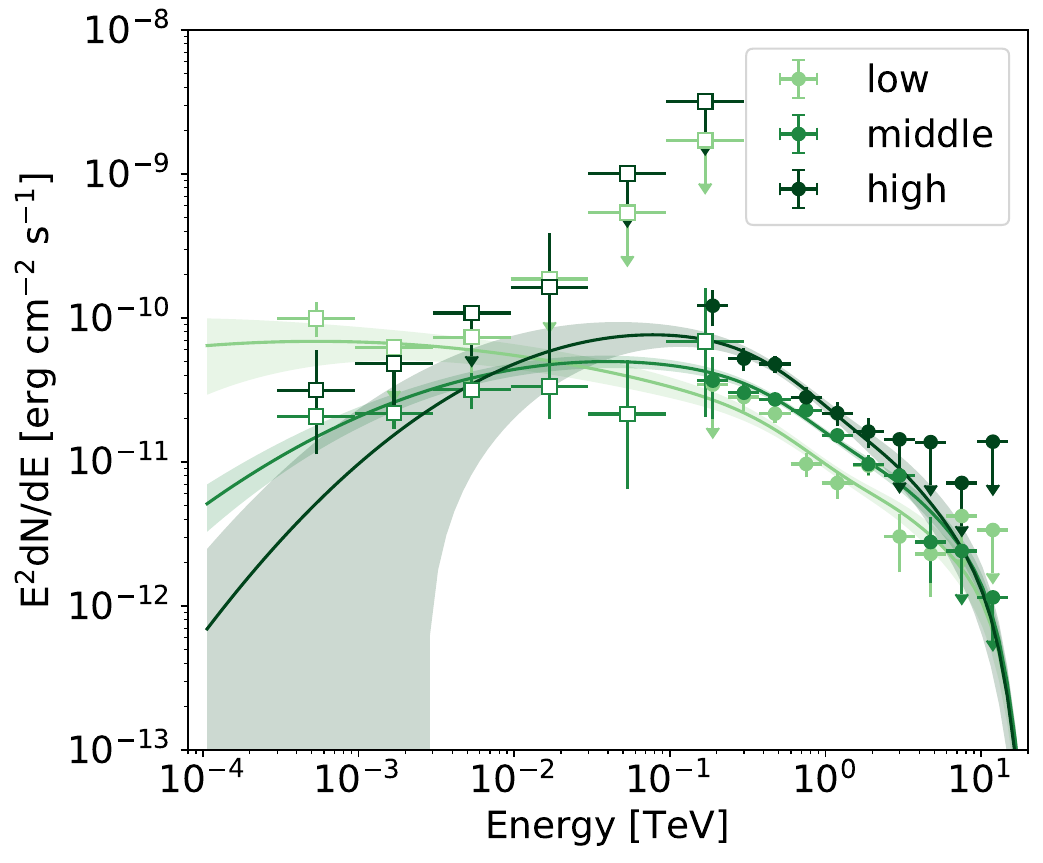}
    \caption{1ES 1959+650}
    \label{fig:1ES1959_joint}
\end{subfigure}
\begin{subfigure}[t]{0.3\textwidth}
\centering
    \includegraphics[width=\textwidth]{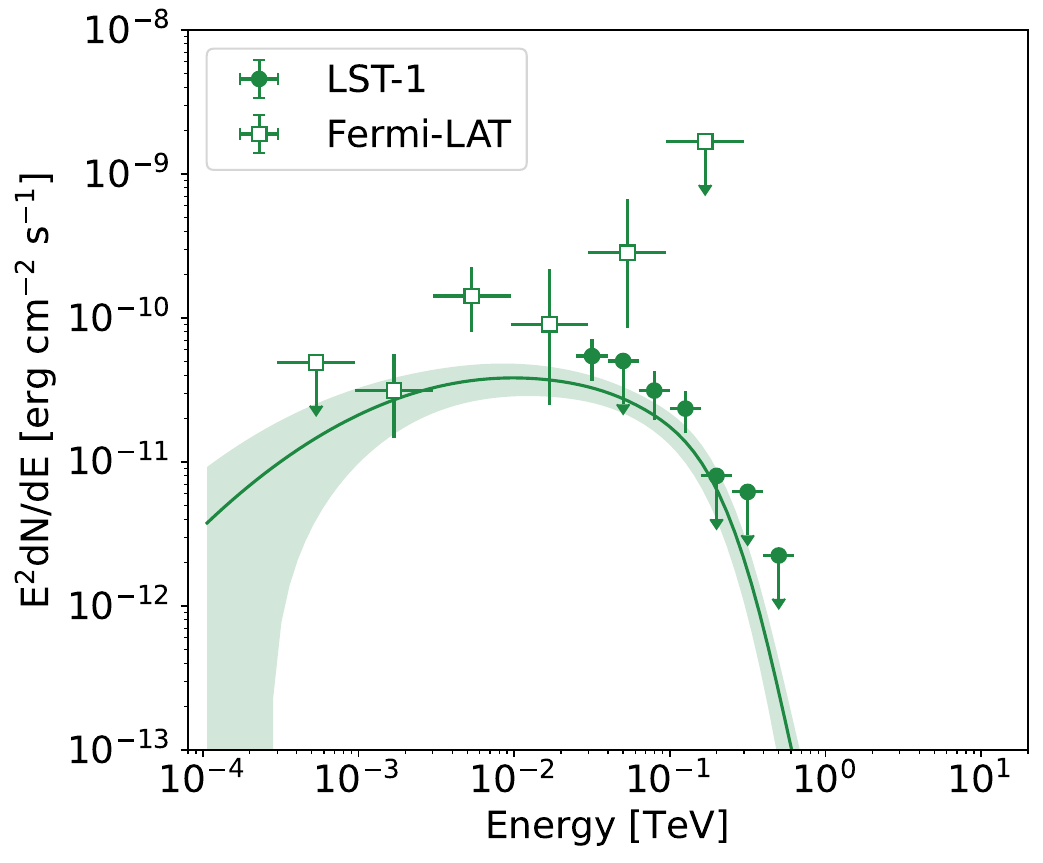}
    \caption{1ES 0647+250}
\end{subfigure}
\begin{subfigure}[t]{0.3\textwidth}
\centering
\includegraphics[width=\textwidth]{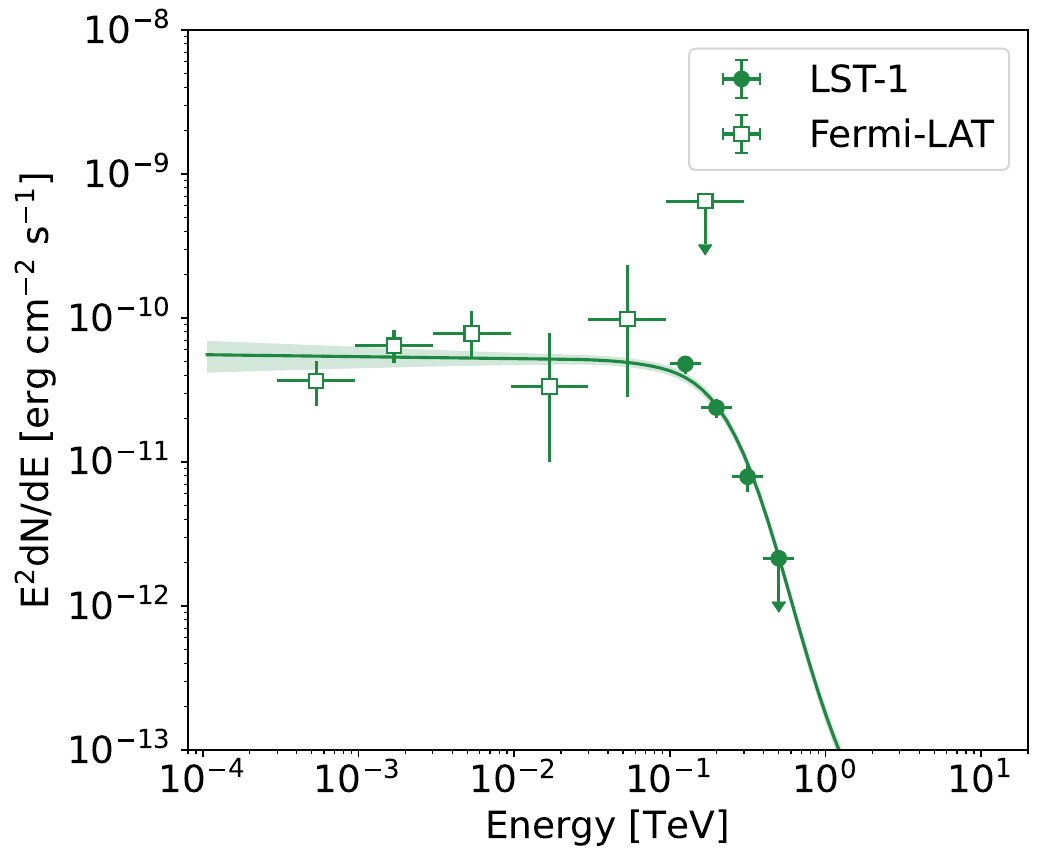}
\caption{PG 1553+113. PWL was used as the model since the LP fits are not well constrained by the data.} 
\label{fig:PG1553_joint}
\end{subfigure}

 \caption{Joint-fit SED of LST-1 (filled circle) and {\it Fermi}-LAT (open square) using a LP model with EBL included. The shaded area shows a statistical error in the spectrum fit.}
 \label{fig:joint_fits}
\end{figure*}

In addition to the spectrum analyses for each instrument, we performed a joint fit analysis with LST-1 and {\it Fermi}-LAT data as shown for all sources in Fig.~\ref{fig:joint_fits}.
For example, Fig.~\ref{fig:1ES1959_joint} shows the SED of 1ES 1959+650.
As can be seen in the figure, we obtained clearer $\gamma$-ray spectral models without gaps for the three spectral states.
This indicates that the joint-fit analysis method can reveal $\gamma$-ray spectra even when {\it Fermi}-LAT data have low statistics in the analysis time period.

The use of a joint dataset has the potential to set much more stringent constraints on the spectral models and similarly differentiate more efficiently between slightly different states.
Note that there are several drawbacks, such as cross-instrument calibrations, artifacts introduced by the different good time intervals (GTIs) in the two instruments, and the use of too simplistic phenomenological models to reproduce the observed broadband emission.
These effects could introduce potential systematics that need to be investigated for each dataset.

Table~\ref{tab:joint} shows spectral parameters of the joint-fit spectrum analysis. 
The reference energy $E_0$ is fixed at 300\, GeV for all the datasets.

\begin{figure}
\centering
\includegraphics[width=0.4\textwidth]{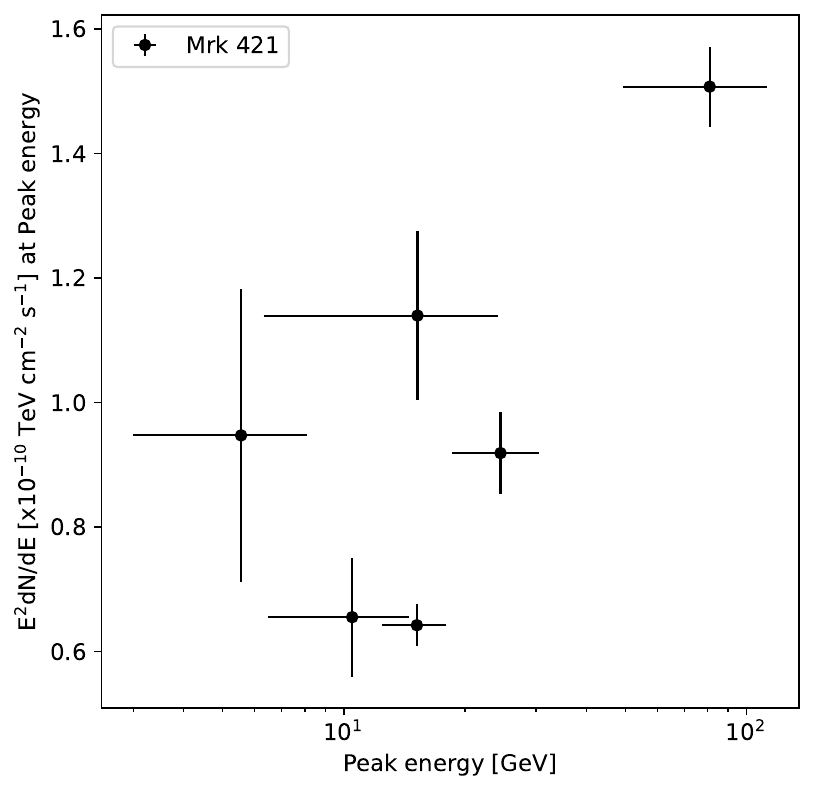}
\caption{Relation between Peak energy and E$^{2}$dN/dE for Mrk 421 in our sample using joint-fit with LP models including EBL.}
\label{fig:sources_eflux_peak}
\end{figure}

The joined fit allows us to constrain the peak of the high-energy contribution of the different blazars and study its evolution. Fig.~\ref{fig:sources_eflux_peak} shows the relation between the energy of the peak and its amplitude for Mrk\,421. It depicts an indication that the peak shifts to higher energies for higher flux levels. 
The determined peak values of all other sources are shown in Table~\ref{tab:joint_peak_lp}. For Mrk\,501, we see three states with very high peak values distributed over different flux values. However, the errors for the determination of the peak are large due to the low curvatures of the spectra. 1ES\, 1959+650 shows a rather low peak energy during its low state, while it shifts to higher values during its higher flux states, which depict similar peak amplitudes. In general, the peak energy of all states and sources lies around the GeV to tens of GeV regime.

\begin{table*}
  \caption{EBL-corrected spectral fit parameters of joint-fit analysis. Statistical errors are described. The reference energy $E_0$ is fixed at 300\,GeV. LRT of the last column shows the preference of fit models of LP over PWL.}
  \label{tab:joint}
  \centering
  \begin{tabular}{llccccccccc}
    \hline \hline
    Source & State & PWL fit & & & LP fit & & & & LRT\\
    & & $f_0 \times 10^{-10}$& $\alpha$ & $\chi^2$/dof & $f_0 \times 10^{-10}$& $\alpha$ & $\beta$ & $\chi^2$/dof & [$\sigma$]\\
    & & [TeV$^{-1}$ cm$^{-2}$ s$^{-1}$] & & & [TeV$^{-1}$ cm$^{-2}$ s$^{-1}$] & & & & \\
    \hline
    Mrk 421 & low & 1.40 $\pm$ 0.07 & 2.36 $\pm$ 0.02 & 172/40 & 2.20 $\pm$ 0.10 & 2.78 $\pm$ 0.05 & 0.10 $\pm$ 0.01 & 49.0/39 & 11.1 \\
     & mid-low & 1.98 $\pm$ 0.08 & 2.25 $\pm$ 0.01 & 159/40 & 2.90 $\pm$ 0.11 & 2.55 $\pm$ 0.03 & 0.08 $\pm$ 0.01 & 28.2/39 & 11.4 \\
     & middle & 3.01 $\pm$ 0.06 & 2.08 $\pm$ 0.01 & 492/40 & 4.23 $\pm$ 0.09 & 2.35 $\pm$ 0.01 & 0.06 $\pm$ 0.01 & 86.1/39 & 20.2 \\
     & high & 4.68 $\pm$ 0.11 & 2.20 $\pm$ 0.01 & 239/40 & 6.11 $\pm$ 0.14 & 2.41 $\pm$ 0.02 & 0.08 $\pm$ 0.01 & 38.3/39 & 14.2 \\
     & post-flare & 7.13 $\pm$ 0.27 & 2.11 $\pm$ 0.02 & 67.3/40 & 8.51 $\pm$ 0.34 & 2.27 $\pm$ 0.03 & 0.04 $\pm$ 0.01 & 36.9/39 & 5.5 \\
     & flare & 14.4 $\pm$ 0.35 & 2.10 $\pm$ 0.02 & 34.3/40 & 15.5 $\pm$ 0.41 & 2.11 $\pm$ 0.02 & 0.04 $\pm$ 0.01 & 16.7/39 & 4.2 \\
    \hline
    Mrk 501 & low & 0.71 $\pm$ 0.06 & 2.26 $\pm$ 0.03 & 51.9/40 & 0.91 $\pm$ 0.09 & 2.38 $\pm$ 0.06 & 0.04 $\pm$ 0.01 & 39.9/39 & 3.5 \\
     & mid-low 1 & 0.91 $\pm$ 0.07 & 2.13 $\pm$ 0.03 & 16.5/40 & 1.09 $\pm$ 0.10 & 2.14 $\pm$ 0.07 & 0.05 $\pm$ 0.03 & 15.4/39 & 1.0 \\
     & mid-low 2  & 1.07 $\pm$ 0.07 & 2.18 $\pm$ 0.02 & 76.0/40 & 1.34 $\pm$ 0.11 & 2.30 $\pm$ 0.04 & 0.04 $\pm$ 0.01 & 34.4/39 & 6.4 \\
     & middle & 1.71 $\pm$ 0.07 & 2.10 $\pm$ 0.02 & 78.0/40 & 2.05 $\pm$ 0.10 & 2.17 $\pm$ 0.03 & 0.04 $\pm$ 0.01 & 55.1/39 & 4.8 \\
     & high 1 & 2.93 $\pm$ 0.09 & 2.16 $\pm$ 0.02 & 52.5/40 & 3.46 $\pm$ 0.12 & 2.26 $\pm$ 0.02 & 0.04 $\pm$ 0.01 & 41.9/39 & 3.3 \\
     & high 2 & 2.97 $\pm$ 0.29 & 2.01 $\pm$ 0.05 & 31.3/40 & 3.21 $\pm$ 0.36 & 2.03 $\pm$ 0.06 & 0.02 $\pm$ 0.02 & 25.1/39 & 2.5 \\
    \hline
    1ES 1959+650 & low & 1.38 $\pm$ 0.11 & 2.24 $\pm$ 0.03 & 23.6/35 & 1.67 $\pm$ 0.20 & 2.33 $\pm$ 0.07 & 0.03 $\pm$ 0.01 & 17.8/34 & 2.4 \\
     & middle & 1.69 $\pm$ 0.09 & 2.03 $\pm$ 0.02 & 88.1/35 & 2.67 $\pm$ 0.18 & 2.26 $\pm$ 0.04 & 0.06 $\pm$ 0.01 & 32.5/34 & 7.4 \\
     & high & 3.05 $\pm$ 0.27 & 2.09 $\pm$ 0.04 & 27.5/35 & 4.44 $\pm$ 0.42 & 2.28 $\pm$ 0.09 & 0.11 $\pm$ 0.04 & 6.83/34 & 4.5 \\
    \hline
    1ES 0647+250 & all & 1.39 $\pm$ 0.38 & 2.06 $\pm$ 0.09 & 19.2/37 & 0.71 $\pm$ 0.35 & 2.77 $\pm$ 0.36 & 0.11 $\pm$ 0.06 & 7.85/36 & 3.4 \\
    \hline
    PG 1553+113 & all & 3.45 $\pm$ 0.32 & 2.01 $\pm$ 0.04 & 19.5/28 & 0.46 $\pm$ 0.08 & 3.67 $\pm$ 0.30 & 0.27 $\pm$ 0.07 & 13.5/27 & 2.4 \\
    \hline
    \end{tabular}
\end{table*}

\begin{table}
  \caption{Peak energy parameters of joint-fit analysis with LP fits including EBL. Statistical errors are described.}
  \label{tab:joint_peak_lp}
  \centering
  \begin{tabular}{llcc}
    \hline \hline
    Source & State & Peak energy & E$^{2}$dN/dE at peak energy \\
    & & [GeV] & [$\times 10^{-11}$ TeV cm$^{-2}$ s$^{-1}$] \\
    \hline
    Mrk 421 & low & 5.55 $\pm$ 2.56 & 9.47 $\pm$ 2.36 \\
     & mid-low & 10.5 $\pm$ 4.0 & 6.55 $\pm$ 0.95 \\
     & middle & 15.2 $\pm$ 2.7 & 6.42 $\pm$ 0.34 \\
     & high & 24.5 $\pm$ 6.0 & 9.19 $\pm$ 0.66 \\
     & post-flare & 15.2 $\pm$ 8.9 & 11.4 $\pm$ 1.4 \\
     & flare & 81.0 $\pm$ 31.6 & 15.1 $\pm$ 0.6 \\
    \hline
    Mrk 501 & low & 2.58 $\pm$ 4.72 & 2.20 $\pm$ 0.90 \\
     & mid-low 1 & 83.5 $\pm$ 80.7 & 1.07 $\pm$ 0.15 \\
     & mid-low 2  & 5.20 $\pm$ 7.36 & 2.23 $\pm$ 0.64 \\
     & middle & 40.4 $\pm$ 23.6 & 2.20 $\pm$ 0.20 \\
     & high 1 & 12.6 $\pm$ 7.1 & 4.71 $\pm$ 0.48 \\
     & high 2 & 154 $\pm$ 246 & 2.92 $\pm$ 0.35 \\
    \hline
    1ES 1959+650 & low & 0.53 $\pm$ 2.02 & 4.29 $\pm$ 3.15 \\
     & middle & 40.7 $\pm$ 17.4 & 3.10 $\pm$ 0.34 \\
     & high & 81.9 $\pm$ 52.9 & 4.78 $\pm$ 0.81 \\
    \hline
    1ES 0647+250 & all & 9.84 $\pm$ 22.7 & 2.39 $\pm$ 3.51 \\
    \hline
    PG 1553+113 & all & 13.7 $\pm$ 13.6 & 5.46 $\pm$ 6.44 \\
    \hline
    \end{tabular}
\end{table}

\section{Potential of LST-1 for extragalactic science}
The number of extragalactic sources emitting VHE $\gamma$ rays detected by current IACTs has been steadily increasing over time.
Recent figures show a rise from 36 (31 blazars) sources in 2009 to 98 (84 blazars) in early 2025, according to TeVCat \citep{2008ICRC....3.1341W}.
Strong absorption of VHE $\gamma$ rays by the EBL has imposed a significant limit on the observational Universe, restricting observations to a maximum redshift of $z \sim 1$.
Despite two decades of operation by leading IACTs such as MAGIC, H.E.S.S., VERITAS, and LST-1, the highest redshift achieved remains $z = 0.997$ recorded for the first time by LST-1 \citep{cortina2023first}, underscoring the persistent challenge of overcoming this barrier.
Given that FSRQs are known to be predominantly located at higher redshifts compared to BL Lacs \citep{Garofalo_2019, Ajello_2022}, it becomes evident that expanding the radius of the observable universe in the VHE $\gamma$-ray domain is crucial to secure a more comprehensive sample of blazars. This expansion is not merely about increasing the number of observed blazars; it also opens up the possibility of capturing detailed phenomena associated with these objects. For instance, blazars are known to exhibit variability on minute timescales \citep{Aharonian_2007, 2019A&A...627A.159H}, and precise monitoring of such rapid fluctuations can yield crucial insights into their emission mechanisms. 

To address the aforementioned challenges in the VHE $\gamma$-ray field, numerous simulation studies have been conducted—and are ongoing—within the CTAO collaboration to explore how CTAO will overcome these limitations.
With its anticipated capabilities and significantly improved sensitivity at the lowest energies, CTAO is expected to expand the reach of VHE blazar observations deeper into the observable universe, up to redshifts of $z \sim 2$ \citep{Abdalla_2021}, and accurately track their flux variability on sub-minute timescales \citep{cerruti2023bright}. However, while these studies have explored the future potential of CTAO, there has been a noticeable lack of quantitative assessments regarding the contributions of LST-1.

In this section, we summarize the findings of a simulation conducted using version 1.2 of {\tt Gammapy}\footnote{\url{https://github.com/gammapy/gammapy/tree/v1.2}} \citep{gammapy:2023} based on the IRFs of LST-1.
The simulation, which we will refer to as the \textit{Blazar Detectability Simulation}, assesses the capability of LST-1 to detect VHE $\gamma$ rays from various redshifted blazars for a given observation time. 

\subsection{Setup for the Blazar Detectability Simulation}
For the \textit{Blazar Detectability Simulation}, we utilized the GeV and TeV flare samples as described in \citet{Abdalla_2021} as input. 
The flare samples were created under the scope of CTAO's AGN-flare program outlined in \citet{Abdalla_2021}, which aims to detect and follow up on VHE flares from AGN, triggered either by external facilities or internally by CTAO's monitoring program. The TeV sample includes 14 AGN observed during flaring states by ground-based VHE instruments, with intrinsic spectra derived by fitting spectral models that account for EBL absorption. The GeV sample was selected from the {\it Fermi}-LAT monitored source list\footnote{\url{https://fermi.gsfc.nasa.gov/ssc/data/access/lat/msl_lc/}}, comprising 63 flares from 63 different AGN, with criteria based on the sources' flux above 1\,GeV and a test statistics $TS \geq 100$. These samples provided a robust foundation for our simulation inputs, allowing us to base our modelled scenarios on actual observed elevated-flux states. 

We accounted for redshift-dependent absorption by the EBL, using the up-to-date model proposed by \citet{2021MNRAS.507.5144S}. \citet{Abdalla_2021} used upper limits for the redshifts of 3C 66A and S5 0716+714 due to the lack of firm values at the time. However, recent updates in the literature have provided more accurate redshift values for these AGN. For 3C 66A, we used the redshift value of $z=0.34$, as reported by \citet{2018MNRAS.474.3162T} following the detection of its host galaxy cluster. Regarding S5 0716+714, based on the archival COS FUV spectra and the identification of its highest-redshift Ly $\alpha$ absorption line as discussed by \citet{2022MNRAS.509.4330D}, we adopted $z=0.26$ for our simulations.

\begin{figure}
    \centering
    \includegraphics[width=0.9\linewidth]{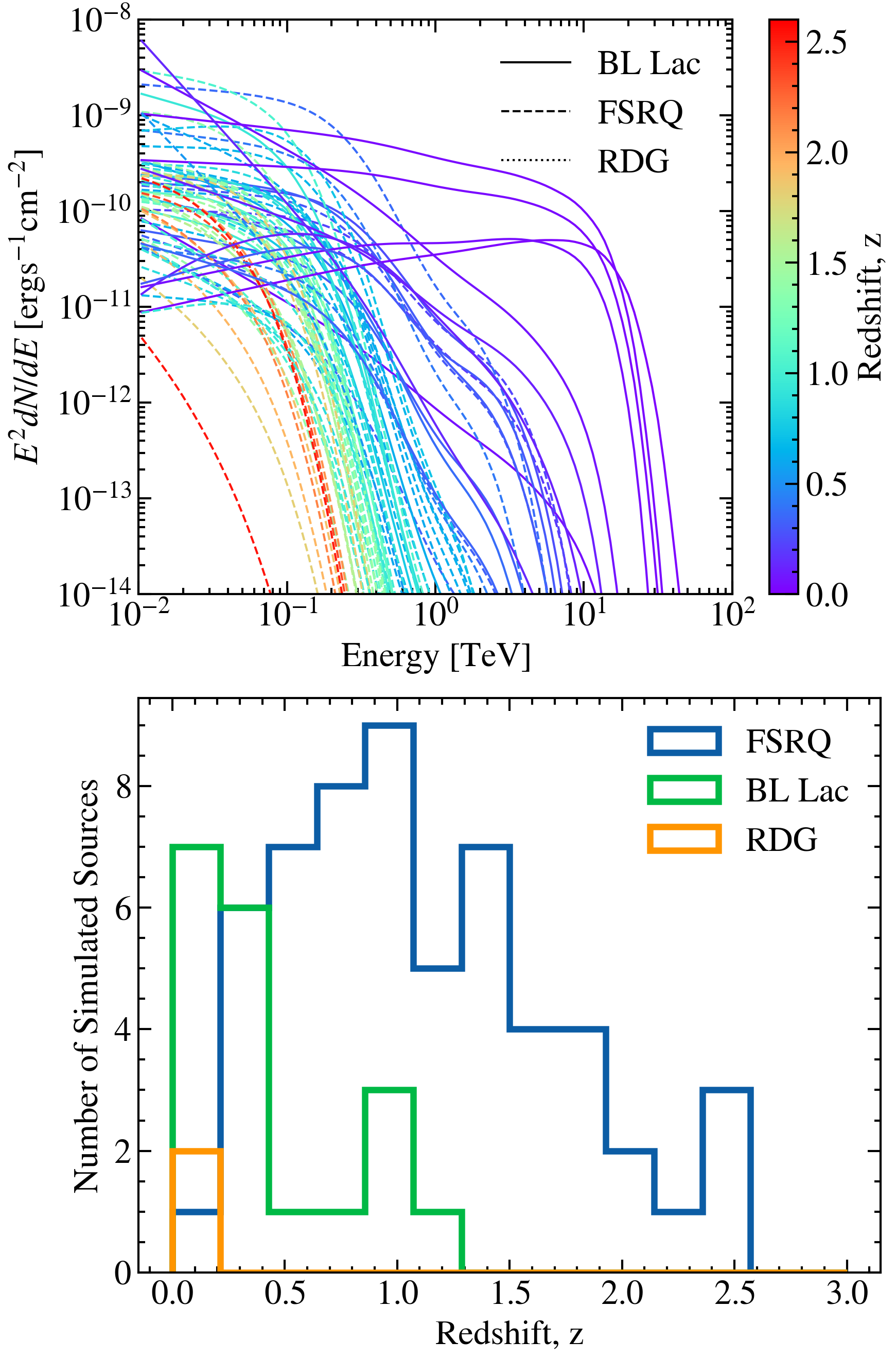}
    \caption{Simulation inputs for the \textit{Blazar Detectability Simulation}. \textit{Top panel}: Spectral models of the blazars used as inputs in the simulation. The spectral model is plotted after accounting for absorption due to the EBL according to each source’s redshift. Labels correspond to source classes, with “RDG” indicating Radio Galaxies. \textit{Bottom panel}: Redshift distribution of the simulated sources by blazar type.}
    \label{fig:simulation_input}
\end{figure}

The intrinsic spectral models were modified to incorporate an exponential cutoff. While the provided GeV and TeV samples were originally modeled using PWL and LP models, simply extrapolating these models without adjustments would be overly optimistic. To address this, following \citet{Abdalla_2021}, we added an exponential cutoff at $1.0 (1+z) \, \mathrm{TeV}$ to the models. We applied these exponential cutoffs to the input PWL and LP models in our simulations. These inputs, including the spectral models and redshift distributions of the blazars used in our simulations, are illustrated in Fig.~\ref{fig:simulation_input}.

Because the IACT energy threshold increases with air mass, we evaluated source visibility with \texttt{astroplan} \citep{2018AJ....155..128M} for December 31, 2025, to December 31, 2026. We retained only sources that accumulate at least 10\,h in astronomical darkness (Sun $< -18^{\circ}$), with Moon illumination $\leq 50\%$ and altitude between $20^{\circ}$ and $85^{\circ}$ at the 4 LSTs site; all others were excluded. For each retained source, we adopt the zenith distance at culmination within that year as the representative value. Observations at larger zenith distance would imply a higher energy threshold; pointings below $20^{\circ}$ require dedicated LST studies \citep{acciari2020magic}.

Once the representative ZD value for each source was determined, we simulated the detector response using the corresponding IRFs for LST-1. 
The IRFs used to mimic the instrument response were selected or constructed as follows: 
We took the background data from the OFF region from Mrk 421 observations and scaled it to the assumed observation time (10\,hrs) to create the mocked Background counts.
If the source's representative ZD was between 0$^\circ$ and 30$^\circ$ (or 30$^\circ$ and 45$^\circ$, above 45$^\circ$), the data of the corresponding ZD range were used.

Lastly, we assumed an observation time of 10\,hrs for all sources and considered that 
LST-1 conducted wobble observations with the source positioned at $0.4 ^\circ$ off-axis from the camera center.

\subsection{Results of the Blazar Detectability Simulation}
Based on the results of the simulations, the detection significance for each source was calculated using the Li \& Ma formula; Eq. (17) in  \citet{li1983analysis}. In this simulation, we computed both the detection significance within the energy range exceeding the $\gamma$-ray horizon ($\tau = 1$) as predicted by \citet{2021MNRAS.507.5144S}, denoted as $\sigma_{\mathrm{Li \& Ma}} ({E > E_{\tau=1}(z)})$, as well as the detection significance across the entire detected energy range, denoted as $\sigma_{\mathrm{Li \& Ma}} (E > E_{\mathrm{min}}^{\mathrm{reco}})$.

\begin{figure}
    \centering
    \includegraphics[width=1\linewidth]{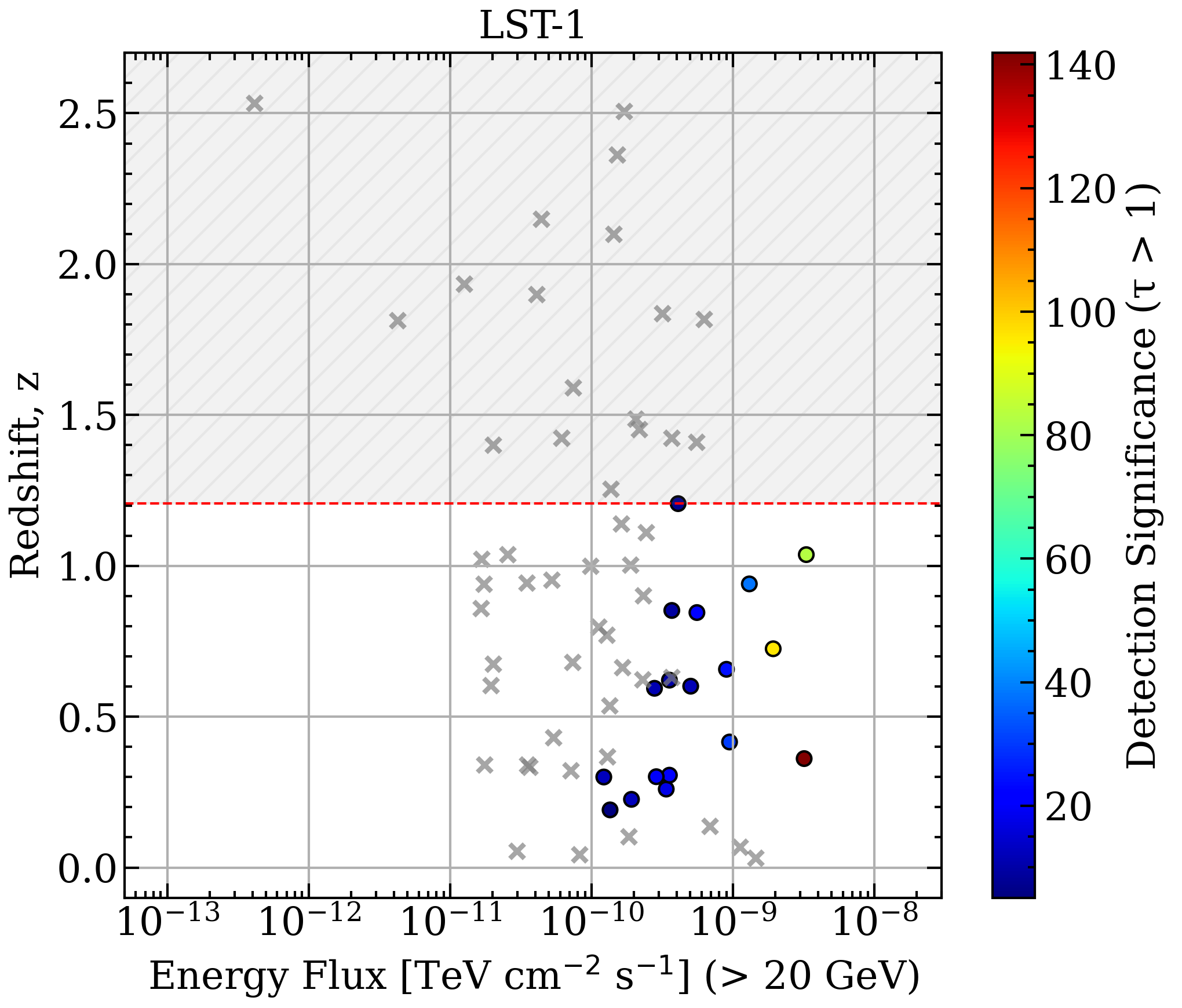}
    \caption{AGN detectability simulation results with a $1.0(1+z)$\,TeV cutoff in all intrinsic spectrum. The color corresponds to detection significance with $\tau > 1$. The x-axis represents the integrated energy flux above 20 GeV. Horizontal and vertical axes represent energy flux and source redshift, respectively. The red dashed line shows the most distant detectable source found in this simulation. The gray shaded area shows the region where AGN cannot be detected with LST-1. 
    }
    \label{fig:simulation_result_1}
\end{figure}

\begin{figure}
    \centering
    \includegraphics[width=1\linewidth]{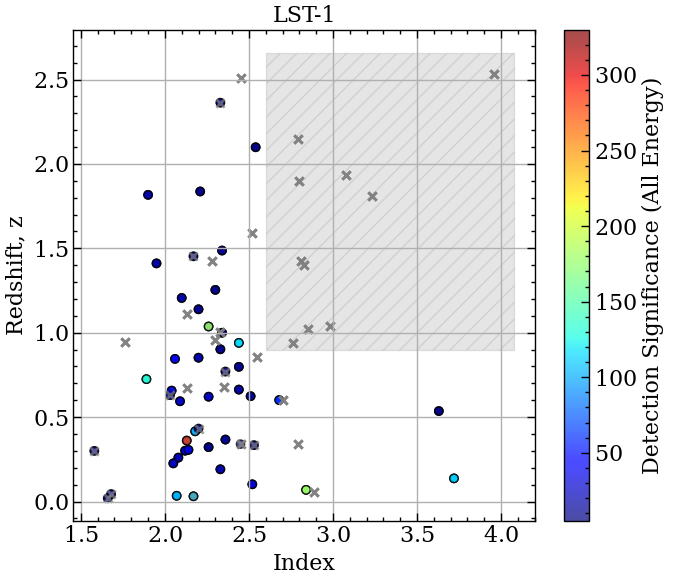}
    \caption{Same as Fig.~\ref{fig:simulation_result_1}, but for detection significance with all energies. The x-axis shows the photon index before EBL absorption, as used as input to the simulations. The horizontal axis represents the spectral index. The Gray shaded area corresponds to the region that will not be detectable by LST-1.
    }
    \label{fig:simulation_result_2}
\end{figure}

The simulation results for the scenario with a 1\,TeV cutoff in all intrinsic spectra are shown in Fig.~\ref{fig:simulation_result_1} and \ref{fig:simulation_result_2}. Fig.~\ref{fig:simulation_result_1} suggests that LST-1 is capable of detecting sources beyond the $\gamma$-ray horizon with a significance greater than 5$\sigma$ up to redshift of approximately $z = 1.2$. 
It corresponds to a flare observed in 2012 from the FSRQ 4C+28.07 (z = 1.206), during which the source reached a $\gamma$-ray flux of 1.6 × 10$^{-6}$\,1/cm$^2$/s in the GeV band. Although this source has not yet been detected by current generation IACTs, its history of multiple flaring events suggests a strong potential for detection by LST-1, with expected significance of $\sigma_{\mathrm{Li \& Ma}} ({E > E_{\tau=1}(z)}) =  7.2\, \sigma$ and $\sigma_{\mathrm{Li \& Ma}} (E > E_{\mathrm{min}}^{\mathrm{reco}}) = 24\, \sigma$, respectively.  
Observing more distant sources in the VHE band with the more sensitive array of four LSTs currently under construction, within dedicated multiwavelength campaigns, could provide important constraints on emission models and offer deeper insights into the physical processes at play.

\section{Discussion and Conclusions}\label{sec:Discussion}

In this work, we present the gamma-ray variability of five bright blazars (Mrk 421, Mrk 501, 1ES 1959+650, 1ES 0647+250, PG 1553+113) using the LST-1 commissioning data accumulated from 2020 to 2022 in combination with \textit{Fermi}-LAT data.
The most variable source of our sample during this time period is Mrk\,421, which also depicts a rapid flare detected on May 18, 2022, with clear intra-night variability. 
The flare of Mrk 421 revealed variability timescales as short as 5\,min, setting limits to the emission size of $(1-5) \times 10^{14}$\,cm, which corresponds to ($3.4 - 17.5$) times gravitational radius. Variability on time scales below 1 hour has already been observed in previous gamma-ray flares of Mrk\,421 \citep{2025A&A...694A.195A,2010A&A...519A..32A}, however, variability down to the order of a few minutes has so far only been seen in lower-energy wavebands, such as X-rays \citep{2024MNRAS.529.1450G}.

\begin{figure}
\centering
\includegraphics[width=0.45\textwidth]{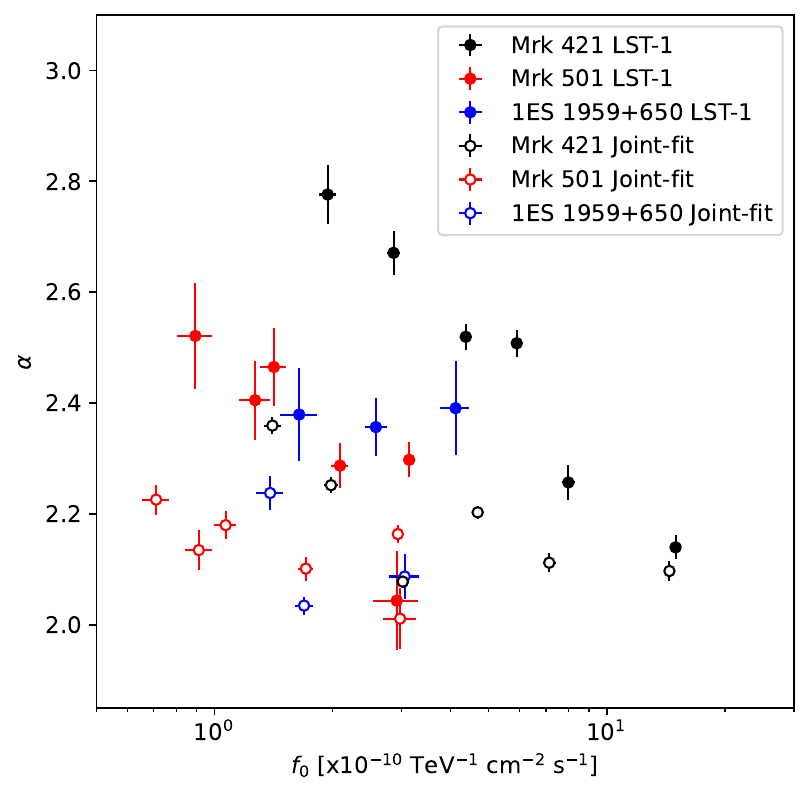}
\caption{Relation between spectral index $\alpha$ and amplitude $f_o$ for the different states of all variable sources in our sample using PWL fits including EBL.}
\label{fig:sources_flux_index}
\end{figure}

The collected sample allows us to compare different gamma-ray flux states across different blazars. Fig.~\ref{fig:sources_flux_index} depicts a comparison between the intrinsic spectral index and the amplitude of the variable sources in our sample. To enable a fair comparison, the PWL results are displayed and discussed here even though log-parabola models are preferred for some of them. This is due to the indirect proportionality between spectral index and curvature in log parabola models. Mrk\,421 is the only source for which a clear harder-when-brighter trend is seen, and especially in the LST-only fits, it tends towards softer indices than Mrk\,501 and 1ES\,1959+650.

All indices of the joint fit results stay between 2.0 to 2.4. Assuming Synchrotron Self Compton emission as the underlying radiation mechanism, we can calculate the spectral index of the relativistic electrons producing the gamma-rays via $p=2\alpha-1$. This results in electron indices between 3 and 3.8 using the simplified assumption that we are in the Thomson regime. Since we expect a broken power-law distribution for the electrons, which steepens by $\Delta p=1$ above its break, these values would be expected from electron distributions with indices ~2.0-2.8. This is in line with shock acceleration as the dominating acceleration mechanism behind these sources \citep{2001ApJ...561..111G}.

Additionally, we studied the variability of the gamma-ray peak position of our sample (see Fig.~\ref{fig:sources_eflux_peak} and Table~\ref{tab:joint_peak_lp}). Mrk\,421 shows indications of its gamma-ray peak shifting to higher energies while increasing in amplitude. This can be explained if the maximum energy of the electron distribution increases, which is expected if the flares were driven by an increase in efficiency of the underlying acceleration mechanisms \citep{1997A&A...320...19M}.

For Mrk\,501, we identify three emission states (mid-low 1, middle, high 2) with rather high peak energies with diverse flux levels. However, taking into account the errors on the peak energies, these high values cannot be claimed significantly. Nonetheless, all three states are adjacent in time, stretching from MJD~59539.07 to 59722.12, 2021 Nov 21 to 2022 May 23. This coincides with a time period in which Mrk\,501 has been depicted as extreme HBL behavior, meaning that its Synchrotron peak frequency shifted to energies > 1\,keV \citep{2024A&A...685A.117M}, which has been explained by changes in the magnetic field or emission region size.

We also simulated the detectability of VHE $\gamma$ rays from various redshifted blazars with LST-1.
Based on the GeV and TeV flare samples as input for our simulation, we calculated the detection significance for each source.
The results suggest that LST-1 is capable of detecting sources beyond the $\gamma$-ray horizon ($\tau = 1$) with a significance greater than 5$\sigma$ up to redshift of approximately $z = 1.2$. 
It corresponds to a flare observed in 2012 from the FSRQ 4C+28.07 ($z = 1.206$), with expected significance of $\sigma_{\mathrm{Li \& Ma}} ({E > E_{\tau=1}(z)}) =  7.2\, \sigma$ and $\sigma_{\mathrm{Li \& Ma}} (E > E_{\mathrm{min}}^{\mathrm{reco}}) = 24\, \sigma$, respectively.
These results foresee an exceptional performance of AGN detection with LST-1 and CTAO in the future, where the full array simulation is ongoing, and the performance will provide information to fully understand the $\gamma$-ray emission mechanism of AGN.

\section*{Acknowledgements}

We gratefully acknowledge financial support from the following agencies and organisations:

Conselho Nacional de Desenvolvimento Cient\'{\i}fico e Tecnol\'{o}gico (CNPq), Funda\c{c}\~{a}o de Amparo \`{a} Pesquisa do Estado do Rio de Janeiro (FAPERJ), Funda\c{c}\~{a}o de Amparo \`{a} Pesquisa do Estado de S\~{a}o Paulo (FAPESP), Funda\c{c}\~{a}o de Apoio \`{a} Ci\^encia, Tecnologia e Inova\c{c}\~{a}o do Paran\'a - Funda\c{c}\~{a}o Arauc\'aria, Ministry of Science, Technology, Innovations and Communications (MCTIC), Brasil;
Ministry of Education and Science, National RI Roadmap Project DO1-153/28.08.2018, Bulgaria;
Croatian Science Foundation (HrZZ) Project IP-2022-10-4595, Rudjer Boskovic Institute, University of Osijek, University of Rijeka, University of Split, Faculty of Electrical Engineering, Mechanical Engineering and Naval Architecture, University of Zagreb, Faculty of Electrical Engineering and Computing, Croatia;
Ministry of Education, Youth and Sports, MEYS  LM2023047, EU/MEYS CZ.02.1.01/0.0/0.0/16\_013/0001403, CZ.02.1.01/0.0/0.0/18\_046/0016007, CZ.02.1.01/0.0/0.0/16\_019/0000754, CZ.02.01.01/00/22\_008/0004632 and CZ.02.01.01/00/23\_015/0008197 Czech Republic;
CNRS-IN2P3, the French Programme d’investissements d’avenir and the Enigmass Labex, 
This work has been done thanks to the facilities offered by the Univ. Savoie Mont Blanc - CNRS/IN2P3 MUST computing center, France;
Max Planck Society, German Bundesministerium f{\"u}r Bildung und Forschung (Verbundforschung / ErUM), Deutsche Forschungsgemeinschaft (SFBs 876 and 1491), Germany;
Istituto Nazionale di Astrofisica (INAF), Istituto Nazionale di Fisica Nucleare (INFN), Italian Ministry for University and Research (MUR), and the financial support from the European Union -- Next Generation EU under the project IR0000012 - CTA+ (CUP C53C22000430006), announcement N.3264 on 28/12/2021: ``Rafforzamento e creazione di IR nell’ambito del Piano Nazionale di Ripresa e Resilienza (PNRR)'';
ICRR, University of Tokyo, JSPS, MEXT, Japan;
JST SPRING - JPMJSP2108;
Narodowe Centrum Nauki, grant number 2023/50/A/ST9/00254, Poland;
The Spanish groups acknowledge the Spanish Ministry of Science and Innovation and the Spanish Research State Agency (AEI) through the government budget lines
PGE2022/28.06.000X.711.04,
28.06.000X.411.01 and 28.06.000X.711.04 of PGE 2023, 2024 and 2025,
and grants PID2019-104114RB-C31,  PID2019-107847RB-C44, PID2019-104114RB-C32, PID2019-105510GB-C31, PID2019-104114RB-C33, PID2019-107847RB-C43, PID2019-107847RB-C42, PID2019-107988GB-C22, PID2021-124581OB-I00, PID2021-125331NB-I00, PID2022-136828NB-C41, PID2022-137810NB-C22, PID2022-138172NB-C41, PID2022-138172NB-C42, PID2022-138172NB-C43, PID2022-139117NB-C41, PID2022-139117NB-C42, PID2022-139117NB-C43, PID2022-139117NB-C44, PID2022-136828NB-C42, PDC2023-145839-I00 funded by the Spanish MCIN/AEI/10.13039/501100011033 and “and by ERDF/EU and NextGenerationEU PRTR; the "Centro de Excelencia Severo Ochoa" program through grants no. CEX2019-000920-S, CEX2020-001007-S, CEX2021-001131-S; the "Unidad de Excelencia Mar\'ia de Maeztu" program through grants no. CEX2019-000918-M, CEX2020-001058-M; the "Ram\'on y Cajal" program through grants RYC2021-032991-I  funded by MICIN/AEI/10.13039/501100011033 and the European Union “NextGenerationEU”/PRTR and RYC2020-028639-I; the "Juan de la Cierva-Incorporaci\'on" program through grant no. IJC2019-040315-I and "Juan de la Cierva-formaci\'on"' through grant JDC2022-049705-I. They also acknowledge the "Atracci\'on de Talento" program of Comunidad de Madrid through grant no. 2019-T2/TIC-12900; the project "Tecnolog\'ias avanzadas para la exploraci\'on del universo y sus componentes" (PR47/21 TAU), funded by Comunidad de Madrid, by the Recovery, Transformation and Resilience Plan from the Spanish State, and by NextGenerationEU from the European Union through the Recovery and Resilience Facility; “MAD4SPACE: Desarrollo de tecnolog\'ias habilitadoras para estudios del espacio en la Comunidad de Madrid" (TEC-2024/TEC-182) project funded by Comunidad de Madrid; the La Caixa Banking Foundation, grant no. LCF/BQ/PI21/11830030; Junta de Andaluc\'ia under Plan Complementario de I+D+I (Ref. AST22\_0001) and Plan Andaluz de Investigaci\'on, Desarrollo e Innovaci\'on as research group FQM-322; Project ref. AST22\_00001\_9 with funding from NextGenerationEU funds; the “Ministerio de Ciencia, Innovaci\'on y Universidades”  and its “Plan de Recuperaci\'on, Transformaci\'on y Resiliencia”; “Consejer\'ia de Universidad, Investigaci\'on e Innovaci\'on” of the regional government of Andaluc\'ia and “Consejo Superior de Investigaciones Cient\'ificas”, Grant CNS2023-144504 funded by MICIU/AEI/10.13039/501100011033 and by the European Union NextGenerationEU/PRTR,  the European Union's Recovery and Resilience Facility-Next Generation, in the framework of the General Invitation of the Spanish Government’s public business entity Red.es to participate in talent attraction and retention programmes within Investment 4 of Component 19 of the Recovery, Transformation and Resilience Plan; Junta de Andaluc\'{\i}a under Plan Complementario de I+D+I (Ref. AST22\_00001), Plan Andaluz de Investigaci\'on, Desarrollo e Innovación (Ref. FQM-322). ``Programa Operativo de Crecimiento Inteligente" FEDER 2014-2020 (Ref.~ESFRI-2017-IAC-12), Ministerio de Ciencia e Innovaci\'on, 15\% co-financed by Consejer\'ia de Econom\'ia, Industria, Comercio y Conocimiento del Gobierno de Canarias; the "CERCA" program and the grants 2021SGR00426 and 2021SGR00679, all funded by the Generalitat de Catalunya; and the European Union's NextGenerationEU (PRTR-C17.I1). This research used the computing and storage resources provided by the Port d’Informaci\'o Cient\'ifica (PIC) data center.
State Secretariat for Education, Research and Innovation (SERI) and Swiss National Science Foundation (SNSF), Switzerland;
The research leading to these results has received funding from the European Union's Seventh Framework Programme (FP7/2007-2013) under grant agreements No~262053 and No~317446;
This project is receiving funding from the European Union's Horizon 2020 research and innovation programs under agreement No~676134;
ESCAPE - The European Science Cluster of Astronomy \& Particle Physics ESFRI Research Infrastructures has received funding from the European Union’s Horizon 2020 research and innovation programme under Grant Agreement no. 824064.

R. Takeishi acknowledges support from JSPS KAKENHI Grant Number JP22K14051. J. Baxter acknowledges support from JSPS KAKENHI Grant Number JP24KJ0545.

This work benefited from the support of the project COCOA-NuGETs ANR-23-CE31-0026 of the French National Research Agency (ANR).
\subsection*{Author contribution}

J. Baxter: paper preparation, LST-1 data analysis of all sources, simulation studies.
G. Di Marco: LST-1 data analysis cross-check of 1ES 1959+650.
L. Heckmann: paper preparation, {\it Fermi}-LAT data analysis of all sources, LST-1 data analysis of Mrk 501, spectral modeling of datasets, spectral variability study, theoretical interpretation.
M. Nievas Rosillo: project coordination, paper preparation, data selection cross-check, {\it Fermi}-LAT data analysis of all sources, creation of {\tt asgardpy}, spectral modeling of datasets, theoretical interpretation.
L. Nickel: data selection cross-check, LST-1 data analysis cross-check.
E. Pons: data selection cross-check, LST-1 data analysis cross-check of 1ES 0647+250 and PG 1553+113.
C. Priyadarshi: project coordination, paper preparation, data selection, generation of MC simulation, LST-1 data analysis of all sources, creation of {\tt asgardpy}, joint-fit analysis of all sources, spectral modeling of datasets, spectral variability study.
D. Sanchez: {\it Fermi}-LAT data analysis of all sources, theoretical interpretation.
R. Takeishi: paper preparation, LST-1 data analysis of Mrk 421 and 1ES 1959+650, spectral modeling of datasets, spectral and temporal variability study of Mrk 421 flare, study on systematic uncertainties using Mrk 421 flare data.
M. V\'{a}zquez Acosta: LST-1 data analysis cross-check, theoretical interpretation.

\section*{Data Availability}

The data underlying this article will be shared on reasonable request to the corresponding author.


\bibliographystyle{mnras}
\bibliography{example} 

\bigskip

\small\it{
$^{1}$ { Department of Physics, Tokai University, 4-1-1, Kita-Kaname, Hiratsuka, Kanagawa 259-1292, Japan }\\
$^{2}$ { Institute for Cosmic Ray Research, University of Tokyo, 5-1-5, Kashiwa-no-ha, Kashiwa, Chiba 277-8582, Japan }\\
$^{3}$ { INFN and Università degli Studi di Siena, Dipartimento di Scienze Fisiche, della Terra e dell'Ambiente (DSFTA), Sezione di Fisica, Via Roma 56, 53100 Siena, Italy }\\
$^{4}$ { Université Paris-Saclay, Université Paris Cité, CEA, CNRS, AIM, F-91191 Gif-sur-Yvette Cedex, France }\\
$^{5}$ { FSLAC IRL 2009, CNRS/IAC, La Laguna, Tenerife, Spain }\\
$^{6}$ { Departament de Física Quàntica i Astrofísica, Institut de Ciències del Cosmos, Universitat de Barcelona, IEEC-UB, Martí i Franquès, 1, 08028, Barcelona, Spain }\\
$^{7}$ { Instituto de Astrofísica de Andalucía-CSIC, Glorieta de la Astronomía s/n, 18008, Granada, Spain }\\
$^{8}$ { Department of Astronomy, University of Geneva, Chemin d'Ecogia 16, CH-1290 Versoix, Switzerland }\\
$^{9}$ { INFN Sezione di Napoli, Via Cintia, ed. G, 80126 Napoli, Italy }\\
$^{10}$ { INAF - Osservatorio Astronomico di Roma, Via di Frascati 33, 00040, Monteporzio Catone, Italy }\\
$^{11}$ { Max-Planck-Institut für Physik, Boltzmannstraße 8, 85748 Garching bei München }\\
$^{12}$ { INFN Sezione di Padova and Università degli Studi di Padova, Via Marzolo 8, 35131 Padova, Italy }\\
$^{13}$ { Instituto de Astrofísica de Canarias and Departamento de Astrofísica, Universidad de La Laguna, C. Vía Láctea, s/n, 38205 La Laguna, Santa Cruz de Tenerife, Spain }\\
$^{14}$ { Univ. Savoie Mont Blanc, CNRS, Laboratoire d'Annecy de Physique des Particules - IN2P3, 74000 Annecy, France }\\
$^{15}$ { Universität Hamburg, Institut für Experimentalphysik, Luruper Chaussee 149, 22761 Hamburg, Germany }\\
$^{16}$ { Graduate School of Science, University of Tokyo, 7-3-1 Hongo, Bunkyo-ku, Tokyo 113-0033, Japan }\\
$^{17}$ { IPARCOS-UCM, Instituto de Física de Partículas y del Cosmos, and EMFTEL Department, Universidad Complutense de Madrid, Plaza de Ciencias, 1. Ciudad Universitaria, 28040 Madrid, Spain }\\
$^{18}$ { Faculty of Science and Technology, Universidad del Azuay, Cuenca, Ecuador. }\\
$^{19}$ { Centro Brasileiro de Pesquisas Físicas, Rua Xavier Sigaud 150, RJ 22290-180, Rio de Janeiro, Brazil }\\
$^{20}$ { CIEMAT, Avda. Complutense 40, 28040 Madrid, Spain }\\
$^{21}$ { University of Geneva - Département de physique nucléaire et corpusculaire, 24 Quai Ernest Ansernet, 1211 Genève 4, Switzerland }\\
$^{22}$ { INFN Sezione di Bari and Politecnico di Bari, via Orabona 4, 70124 Bari, Italy }\\
$^{23}$ { Institut de Fisica d'Altes Energies (IFAE), The Barcelona Institute of Science and Technology, Campus UAB, 08193 Bellaterra (Barcelona), Spain }\\
$^{24}$ { INAF - Osservatorio Astronomico di Brera, Via Brera 28, 20121 Milano, Italy }\\
$^{25}$ { Faculty of Physics and Applied Informatics, University of Lodz, ul. Pomorska 149-153, 90-236 Lodz, Poland }\\
$^{26}$ { INAF - Osservatorio di Astrofisica e Scienza dello spazio di Bologna, Via Piero Gobetti 93/3, 40129 Bologna, Italy }\\
$^{27}$ { Dipartimento di Fisica e Astronomia (DIFA) Augusto Righi, Università di Bologna, via Gobetti 93/2, I-40129 Bologna, Italy }\\
$^{28}$ { Lamarr Institute for Machine Learning and Artificial Intelligence, 44227 Dortmund, Germany }\\
$^{29}$ { INFN Sezione di Trieste and Università degli studi di Udine, via delle scienze 206, 33100 Udine, Italy }\\
$^{30}$ { INAF - Istituto di Astrofisica e Planetologia Spaziali (IAPS), Via del Fosso del Cavaliere 100, 00133 Roma, Italy }\\
$^{31}$ { Aix Marseille Univ, CNRS/IN2P3, CPPM, Marseille, France }\\
$^{32}$ { INFN Sezione di Bari and Università di Bari, via Orabona 4, 70126 Bari, Italy }\\
$^{33}$ { INFN Sezione di Torino, Via P. Giuria 1, 10125 Torino, Italy }\\
$^{34}$ { Dipartimento di Fisica - Universitá degli Studi di Torino, Via Pietro Giuria 1 - 10125 Torino, Italy }\\
$^{35}$ { Palacky University Olomouc, Faculty of Science, 17. listopadu 1192/12, 771 46 Olomouc, Czech Republic }\\
$^{36}$ { Dipartimento di Fisica e Chimica 'E. Segrè' Università degli Studi di Palermo, via delle Scienze, 90128 Palermo }\\
$^{37}$ { INFN Sezione di Catania, Via S. Sofia 64, 95123 Catania, Italy }\\
$^{38}$ { IRFU, CEA, Université Paris-Saclay, Bât 141, 91191 Gif-sur-Yvette, France }\\
$^{39}$ { Port d'Informació Científica, Edifici D, Carrer de l'Albareda, 08193 Bellaterrra (Cerdanyola del Vallès), Spain }\\
$^{40}$ { University of Alcalá UAH, Departamento de Physics and Mathematics, Pza. San Diego, 28801, Alcalá de Henares, Madrid, Spain }\\
$^{41}$ { INFN Sezione di Bari, via Orabona 4, 70125, Bari, Italy }\\
$^{42}$ { Department of Physics, TU Dortmund University, Otto-Hahn-Str. 4, 44227 Dortmund, Germany }\\
$^{43}$ { University of Rijeka, Department of Physics, Radmile Matejcic 2, 51000 Rijeka, Croatia }\\
$^{44}$ { Institute for Theoretical Physics and Astrophysics, Universität Würzburg, Campus Hubland Nord, Emil-Fischer-Str. 31, 97074 Würzburg, Germany }\\
$^{45}$ { Department of Physics and Astronomy, University of Turku, Finland, FI-20014 University of Turku, Finland }\\
$^{46}$ { Department of Physics, TU Dortmund University, Otto-Hahn-Str. 4, 44227 Dortmund, Germany }\\
$^{47}$ { INFN Sezione di Roma La Sapienza, P.le Aldo Moro, 2 - 00185 Rome, Italy }\\
$^{48}$ { ILANCE, CNRS – University of Tokyo International Research Laboratory, University of Tokyo, 5-1-5 Kashiwa-no-Ha Kashiwa City, Chiba 277-8582, Japan }\\
$^{49}$ { Physics Program, Graduate School of Advanced Science and Engineering, Hiroshima University, 1-3-1 Kagamiyama, Higashi-Hiroshima City, Hiroshima, 739-8526, Japan }\\
$^{50}$ { INFN Sezione di Roma Tor Vergata, Via della Ricerca Scientifica 1, 00133 Rome, Italy }\\
$^{51}$ { University of Split, FESB, R. Boškovića 32, 21000 Split, Croatia }\\
$^{52}$ { Department of Physics, Yamagata University, 1-4-12 Kojirakawa-machi, Yamagata-shi, 990-8560, Japan }\\
$^{53}$ { Institut für Theoretische Physik, Lehrstuhl IV: Plasma-Astroteilchenphysik, Ruhr-Universität Bochum, Universitätsstraße 150, 44801 Bochum, Germany }\\
$^{54}$ { Sendai College, National Institute of Technology, 4-16-1 Ayashi-Chuo, Aoba-ku, Sendai city, Miyagi 989-3128, Japan }\\
$^{55}$ { Université Paris Cité, CNRS, Astroparticule et Cosmologie, F-75013 Paris, France }\\
$^{56}$ { Josip Juraj Strossmayer University of Osijek, Department of Physics, Trg Ljudevita Gaja 6, 31000 Osijek, Croatia }\\
$^{57}$ { Department of Astronomy and Space Science, Chungnam National University, Daejeon 34134, Republic of Korea }\\
$^{58}$ { INFN Dipartimento di Scienze Fisiche e Chimiche - Università degli Studi dell'Aquila and Gran Sasso Science Institute, Via Vetoio 1, Viale Crispi 7, 67100 L'Aquila, Italy }\\
$^{59}$ { Chiba University, 1-33, Yayoicho, Inage-ku, Chiba-shi, Chiba, 263-8522 Japan }\\
$^{60}$ { Kitashirakawa Oiwakecho, Sakyo Ward, Kyoto, 606-8502, Japan }\\
$^{61}$ { FZU - Institute of Physics of the Czech Academy of Sciences, Na Slovance 1999/2, 182 21 Praha 8, Czech Republic }\\
$^{62}$ { Laboratory for High Energy Physics, École Polytechnique Fédérale, CH-1015 Lausanne, Switzerland }\\
$^{63}$ { Astronomical Institute of the Czech Academy of Sciences, Bocni II 1401 - 14100 Prague, Czech Republic }\\
$^{64}$ { Faculty of Science, Ibaraki University, 2 Chome-1-1 Bunkyo, Mito, Ibaraki 310-0056, Japan }\\
$^{65}$ { Sorbonne Université, CNRS/IN2P3, Laboratoire de Physique Nucléaire et de Hautes Energies, LPNHE, 4 place Jussieu, 75005 Paris, France }\\
$^{66}$ { Graduate School of Science and Engineering, Saitama University, 255 Simo-Ohkubo, Sakura-ku, Saitama city, Saitama 338-8570, Japan }\\
$^{67}$ { Institute of Particle and Nuclear Studies, KEK (High Energy Accelerator Research Organization), 1-1 Oho, Tsukuba, 305-0801, Japan }\\
$^{68}$ { INFN Sezione di Trieste and Università degli Studi di Trieste, Via Valerio 2 I, 34127 Trieste, Italy }\\
$^{69}$ { Escuela Politécnica Superior de Jaén, Universidad de Jaén, Campus Las Lagunillas s/n, Edif. A3, 23071 Jaén, Spain }\\
$^{70}$ { Saha Institute of Nuclear Physics, A CI of Homi Bhabha National
Institute, Kolkata 700064, West Bengal, India }\\
$^{71}$ { Institute for Nuclear Research and Nuclear Energy, Bulgarian Academy of Sciences, 72 boul. Tsarigradsko chaussee, 1784 Sofia, Bulgaria }\\
$^{72}$ { Department of Physics and Astronomy, Clemson University, Kinard Lab of Physics, Clemson, SC 29634, USA }\\
$^{73}$ { Institut de Fisica d'Altes Energies (IFAE), The Barcelona Institute of Science and Technology, Campus UAB, 08193 Bellaterra (Barcelona), Spain }\\
$^{74}$ { Grupo de Electronica, Universidad Complutense de Madrid, Av. Complutense s/n, 28040 Madrid, Spain }\\
$^{75}$ { Macroarea di Scienze MMFFNN, Università di Roma Tor Vergata, Via della Ricerca Scientifica 1, 00133 Rome, Italy }\\
$^{76}$ { Institute of Space Sciences (ICE, CSIC), and Institut d'Estudis Espacials de Catalunya (IEEC), and Institució Catalana de Recerca I Estudis Avançats (ICREA), Campus UAB, Carrer de Can Magrans, s/n 08193 Bellatera, Spain }\\
$^{77}$ { Department of Physics, Konan University, 8-9-1 Okamoto, Higashinada-ku Kobe 658-8501, Japan }\\
$^{78}$ { School of Allied Health Sciences, Kitasato University, Sagamihara, Kanagawa 228-8555, Japan }\\
$^{79}$ { RIKEN, Institute of Physical and Chemical Research, 2-1 Hirosawa, Wako, Saitama, 351-0198, Japan }\\
$^{80}$ { Charles University, Institute of Particle and Nuclear Physics, V Holešovičkách 2, 180 00 Prague 8, Czech Republic }\\
$^{81}$ { Division of Physics and Astronomy, Graduate School of Science, Kyoto University, Sakyo-ku, Kyoto, 606-8502, Japan }\\
$^{82}$ { Institute for Space-Earth Environmental Research, Nagoya University, Chikusa-ku, Nagoya 464-8601, Japan }\\
$^{83}$ { Kobayashi-Maskawa Institute (KMI) for the Origin of Particles and the Universe, Nagoya University, Chikusa-ku, Nagoya 464-8602, Japan }\\
$^{84}$ { Graduate School of Technology, Industrial and Social Sciences, Tokushima University, 2-1 Minamijosanjima,Tokushima, 770-8506, Japan }\\
$^{85}$ { INFN Sezione di Pisa, Edificio C – Polo Fibonacci, Largo Bruno Pontecorvo 3, 56127 Pisa, Italy }\\
$^{86}$ { Gifu University, Faculty of Engineering, 1-1 Yanagido, Gifu 501-1193, Japan }\\
$^{87}$ { Department of Physical Sciences, Aoyama Gakuin University, Fuchinobe, Sagamihara, Kanagawa, 252-5258, Japan }\\}


\appendix

\section{Spectral parameters of each sources}\label{sec:fit_params}
\begin{figure*}
\centering
    \begin{subfigure}[t]{1\textwidth}
        \centering
        \includegraphics[width=0.5\textwidth]{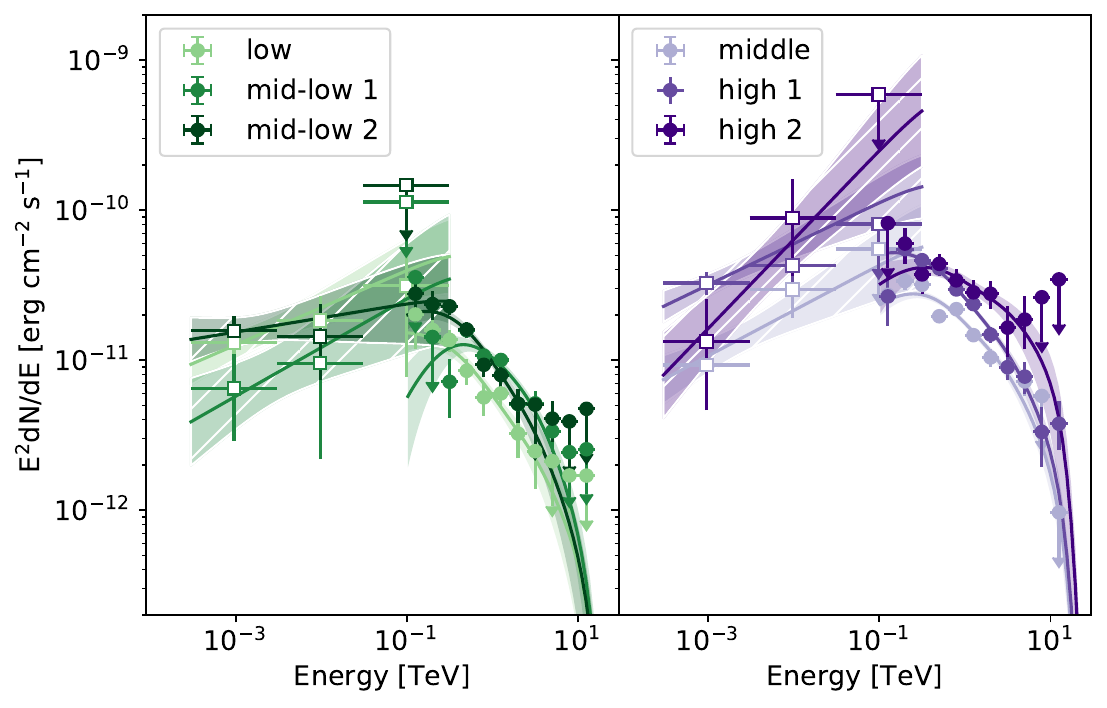}
        \caption{Mrk 501}
    \label{fig:Mrk501_LST_Fermi_SED}
    \end{subfigure}
    \begin{subfigure}[t]{0.32\textwidth}
        \centering
        \includegraphics[width=\textwidth]{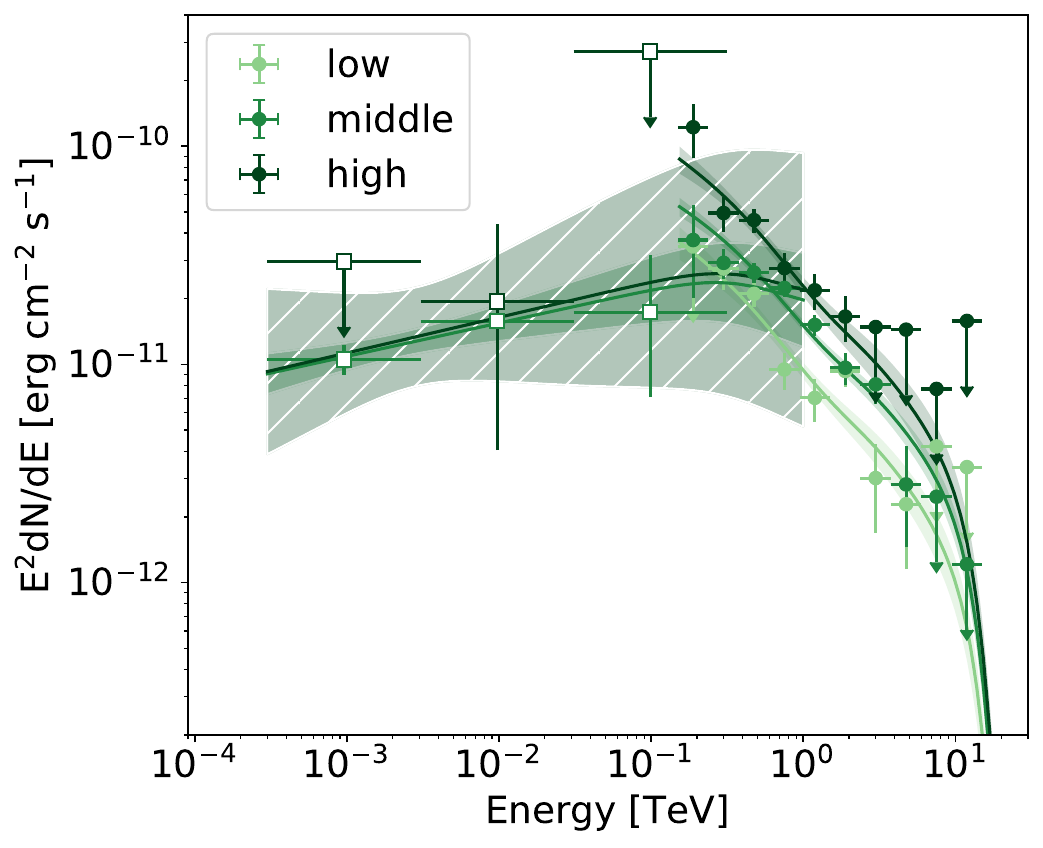}
        \caption{1ES 1959+1650}
        \label{fig:1ES1959+650_LST_Fermi_SED}
    \end{subfigure}
    \hspace{1mm}
    \begin{subfigure}[t]{0.32\textwidth}
        \centering
        \includegraphics[width=\textwidth]{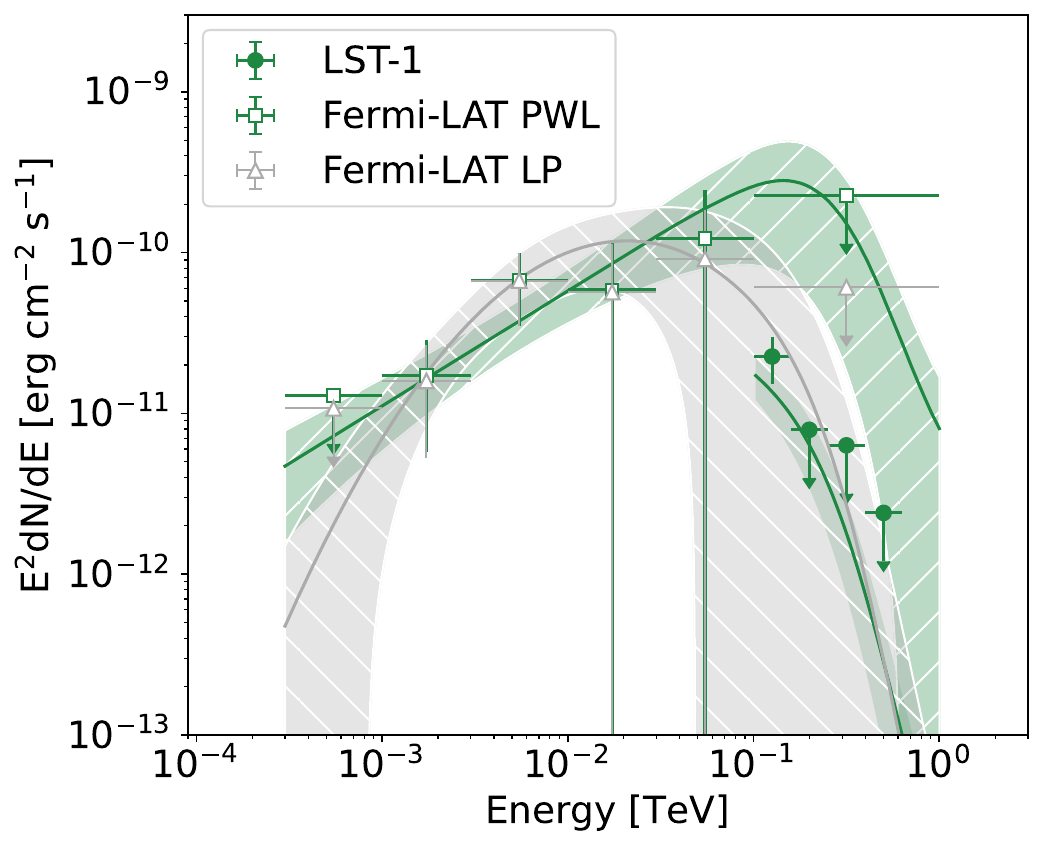}
        \caption{1ES 0647+250 with a PWL fit model used in the \textit{Fermi}-LAT and LST-1 fits (green). Additionally, a LP model fit of the \textit{Fermi}-LAT data is shown (gray).}
    \label{fig:1ES0647+250_LST_Fermi_SED}
    \end{subfigure}
    \hspace{1mm}
    \begin{subfigure}[t]{0.32\textwidth}
        \centering
        \includegraphics[width=\textwidth]{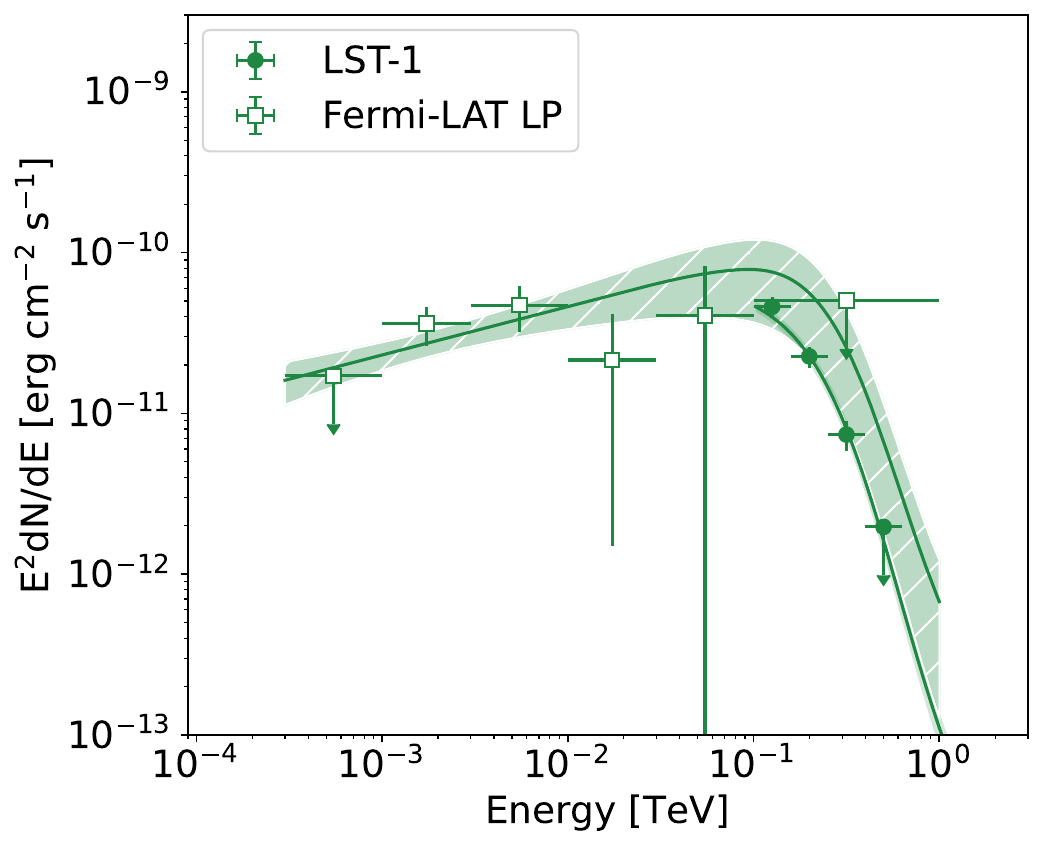}
        \caption{PG 1553+113}
    \label{fig:PG1553+113_LST_Fermi_SED}
    \end{subfigure}
\caption{Same as Fig.~\ref{fig:Mrk421_LST_Fermi_SED}, but for different sources. For Mrk 501, LP and PWL models are used for LST-1 and \textit{Fermi}-LAT data, respectively. A PWL model is used for both the LST-1 and \textit{Fermi}-LAT data for the other three sources. The corresponding model parameters are shown in Table~\ref{tab:Mrk501_LST_Fermi_SED}.}
\label{fig:LST_Fermi_SED}
\end{figure*}

This section presents additional material for the discussion in Section~\ref{subsec:spec_var}.
Table~\ref{tab:Mrk501_LST_Fermi_SED} shows spectral fit parameters of sources except for Mrk 421, which is shown in Table~\ref{tab:Mrk421_LST_Fermi_SED}. Table~\ref{tab:fermi_fit} shows the corresponding parameters of the spectral fits for Fermi-LAT. Fig.~\ref{fig:LST_Fermi_SED} shows the corresponding SEDs.

\begin{table*}
  \caption{Same as Table~\ref{tab:Mrk421_LST_Fermi_SED}, but for other sources. The reference energy $E_0$ is fixed at 300\,GeV. LRT of the last column shows the preference of fit models of LP over PWL. } 
  \label{tab:Mrk501_LST_Fermi_SED}
  \centering
  \begin{tabular}{lccccccccc}
  \hline \hline
    State & Time & PWL fit & & & LP fit & & & & LRT\\
     & [MJD] & $f_0 \times 10^{-10}$& $\alpha$ & $\chi^2$/dof & $f_0 \times 10^{-10}$& $\alpha$ & $\beta$ & $\chi^2$/dof & [$\sigma$]\\
     & & [TeV$^{-1}$ cm$^{-2}$ s$^{-1}$] & & & [TeV$^{-1}$ cm$^{-2}$ s$^{-1}$] & & & &\\
     \hline
    & &  &  &  &  & & &  &\\ 
    Mrk 501 &  &  &  &  &  & & &  &\\
    \hline 
    low & 59357.56-59539.07 & 0.83 $\pm$ 0.08 & 2.38 $\pm$ 0.07 & 33.2/18 & 0.91 $\pm$ 0.10 & 2.26 $\pm$ 0.13 & 0.11 $\pm$ 0.07 & 20.5/17 & 3.5 \\
     mid-low 1 & 59539.07-59628.75 & 1.02 $\pm$ 0.09 & 2.21 $\pm$ 0.05 & 10.2/18 & 0.88 $\pm$ 0.18 & 1.55 $\pm$ 0.33 & 0.27 $\pm$ 0.11 & 6.54/17 & 1.9 \\
     & 59643.21-59647.70 & & & & & & & & \\
     mid-low 2 & 59040.95-59331.11 & 1.24 $\pm$ 0.10 & 2.30 $\pm$ 0.05 & 44.7/18 & 1.45 $\pm$ 0.14 & 2.15 $\pm$ 0.11 & 0.17 $\pm$ 0.07 & 16.0/17 & 5.4 \\
     middle & 59647.70-59722.12 & 1.92 $\pm$ 0.10 & 2.20 $\pm$ 0.34 & 64.8/18 & 2.03 $\pm$ 0.12 & 1.96 $\pm$ 0.08 & 0.15 $\pm$ 0.04 & 38.3/17 & 5.2 \\
     high 1 & 59331.11-59357.56 & 3.29 $\pm$ 0.11 & 2.29 $\pm$ 0.02 & 49.6/18 & 3.53 $\pm$ 0.13 & 2.14 $\pm$ 0.05 & 0.12 $\pm$ 0.03 & 31.7/17 & 4.2 \\
     high 2 & 59628.75-59643.21 & 3.04 $\pm$ 0.35 & 2.03 $\pm$ 0.07 & 14.0/18 & 3.06 $\pm$ 0.41 & 1.83 $\pm$ 0.19 & 0.11 $\pm$ 0.08 & 10.2/17 & 1.9 \\
     \hline
    & &  &  &  &  & & &  &\\ 
    & &  &  &  &  & & &  &\\ 
    1ES 1959+650  & &  &  &  &  & & &  &\\
    \hline
     low & 59183.11-59700.19 & 1.87 $\pm$ 0.22 & 2.48 $\pm$ 0.10 & 14.9/13 & 1.85 $\pm$ 0.24 & 2.44 $\pm$ 0.25 & 0.02 $\pm$ 0.10 & 14.8/12 & 0.2 \\
     middle & 59042.01-59071.48 & 2.81 $\pm$ 0.19 & 2.43 $\pm$ 0.06 & 21.3/13 & 2.45 $\pm$ 0.25 & 1.79 $\pm$ 0.22 & 0.29 $\pm$ 0.10 & 6.46/12 & 3.8 \\
     & 59700.19-59705.17 & & & & & & & & \\
     high & 59071.48-59183.11 & 4.51 $\pm$ 0.41 & 2.47 $\pm$ 0.10 & 6.95/13 & 4.48 $\pm$ 0.43 & 2.33 $\pm$ 0.25 & 0.08 $\pm$ 0.13 & 6.08/12 & 0.9 \\
      \hline
        & &  &  &  &  & & &  &\\ 
    & &  &  &  &  & & &  &\\ 
    
    1ES 0647+250 & &  &  &  &  & & &  &\\
    \hline
    all & - & 0.72 $\pm$ 0.42 & 2.64 $\pm$ 0.30 & 5.86/18 & 0.45 $\pm$ 0.28 & 3.45 $\pm$ 1.11 & 0.26 $\pm$ 0.40 & 5.73/17 & 0.4 \\
     \hline
        & &  &  &  &  & & &  &\\ 
    & &  &  &  &  & & &  &\\
    PG 1553+113 & &  &  &  &  & & &  &\\
    \hline 
    all & - & 2.88 $\pm$ 0.38 & 2.26 $\pm$ 0.10 & 7.20/18 & 1.82 $\pm$ 0.63 & 3.42 $\pm$ 0.65 & 0.46 $\pm$ 0.25 & 1.23/17 & 2.4\\
    \hline
  \end{tabular}
\end{table*}

\begin{table*}
  \caption{Same as Table~\ref{tab:Mrk421_LST_Fermi_SED} and Table~\ref{tab:Mrk501_LST_Fermi_SED}, but for \textit{Fermi}-LAT PWL fit results.}
  \label{tab:fermi_fit}
  \centering
  \begin{tabular}{llcccc}
    \hline \hline
    Source & State & $E_{0}$ & $f_0 \times 10^{-12}$& $\alpha$ & $TS$  \\
    & & [GeV] & [MeV$^{-1}$ cm$^{-2}$ s$^{-1}$] & & \\
    \hline
    Mrk 421 & low & 1.26 & 20.4 $\pm$ 3.0 & 2.04 $\pm$ 0.14 & 206\\
     & mid-low & 1.26 & 13.1 $\pm$ 2.0 & 1.83 $\pm$ 0.12 & 212\\
     & middle & 1.26& 12.9 $\pm$ 0.3 & 1.81 $\pm$ 0.02 & 1.22$\times 10^{4}$\\
     & high & 1.26 & 15.2 $\pm$ 2.6 & 1.77 $\pm$ 0.12 & 205\\
     & post-flare & 1.26 & 28.8 $\pm$ 4.2 & 1.97 $\pm$ 0.13 & 276\\
     & flare & 1.26 & 37.4 $\pm$ 10.0 & 1.91 $\pm$ 0.23 & 87.6\\
    \hline
    Mrk 501 & low & 1.51 & 3.85 $\pm$ 0.84 & 1.75 $\pm$ 0.15 & 125\\
     & mid-low 1 & 1.51 &180 $\pm$ 1.07 & 1.68 $\pm$ 0.34 & 14.9 \\
     & mid-low 2 & 1.51 & 4.40 $\pm$ 1.01  & 1.91 $\pm$ 0.20 & 88.1\\
     & middle & 1.51 & 3.32 $\pm$ 0.90 & 1.70 $\pm$ 0.18 & 75.0\\
     & high 1 &1.51 &9.75 $\pm$ 1.67 & 1.73 $\pm$ 0.12 &203\\
     & high 2 & 1.51 & 5.69 $\pm$ 3.49 & 1.41 $\pm$ 0.29 & 30.5\\
    \hline
    1ES 1959+650 & low & 1.74 & 2.40 $\pm$ 0.71 & 3.00 $\pm$ 6.7$\times 10^{-5}$ & 27.0 \\
     & middle & 1.74 & 2.45 $\pm$ 0.34 & 1.85 $\pm$ 0.10 & 226\\
     & high & 1.74 & 2.56 $\pm$ 1.88 & 1.84 $\pm$ 0.54 & 8.35\\
    \hline
    1ES 0647+250 & all & 1.00 & 6.94 $\pm$ 3.33 & 1.29 $\pm$ 0.20 & 59.0\\
    \hline
    PG 1553+113 & all & 1.00 & 14.4 $\pm$ 2.9 & 1.70 $\pm$ 0.13 & 154\\
    \hline
    \end{tabular}
\end{table*}


\bsp	
\label{lastpage}
\end{document}